\documentclass[twocolumn, tighten, times]{aastex631}

\usepackage{graphicx}
\usepackage{epstopdf}
\usepackage{float} 
\usepackage{bm}
\usepackage{color}
\usepackage{soul}
\usepackage{url}
\usepackage{rotating}  
\usepackage{threeparttablex, tablefootnote} 
\usepackage{enumitem} 
\usepackage{microtype}
\usepackage{tabularx}
\usepackage{booktabs}
\usepackage{makecell}
\shorttitle{Robustness Analysis of USmorph}
\shortauthors{Zhu. ET AL}
\graphicspath{{./}{figures/}}

\begin{document}

\title{Robustness Analysis of USmorph: I. Generalization Efficiency of Unsupervised Strategies and Supervised Learning in Galaxy Morphological Classification}

\author[0009-0004-0966-6439]{Shiwei Zhu}
\affil{School of Mathematics and Physics, Anqing Normal University, Anqing 246011, China; 
\url{wen@mail.ustc.edu.cn}} 
\affil{Institute of Astronomy and Astrophysics, Anqing Normal University, Anqing 246133, China}

\author[0000-0001-9694-2171]{Guanwen Fang}
\altaffiliation{Corresponding author: Guanwen Fang}
\affil{School of Mathematics and Physics, Anqing Normal University, Anqing 246011, China; 
\url{wen@mail.ustc.edu.cn}} 
\affil{Institute of Astronomy and Astrophysics, Anqing Normal University, Anqing 246133, China}

\author[0000-0002-4638-0235]{Yao Dai}
\affil{Shanghai Astronomical Observatory, Chinese Academy of Sciences, 80 Nandan Road, Shanghai 200030, China}
\affil{School of Astronomy and Space Science, University of Chinese Academy of Sciences, No. 19A Yuquan Road, Beijing 100049, China}

\author[0000-0002-5133-2668]{Chichun Zhou}
\affil{School of Engineering, Dali University, Dali 671003, China}

\author[0000-0001-7707-5930]{Yirui Zheng}
\affil{School of Mathematics and Physics, Anqing Normal University, Anqing 246011, China;
\url{wen@mail.ustc.edu.cn}} 
\affil{Institute of Astronomy and Astrophysics, Anqing Normal University, Anqing 246133, China}

\author[0000-0002-0846-7591]{Jie Song}
\affil{Department of Astronomy, University of Science and Technology of China, Hefei 230026, China; \url{xkong@ustc.edu.cn}} 
\affil{School of Astronomy and Space Science, University of Science and Technology of China, Hefei 230026, China}
\affil{Institute of Deep Space Sciences, Deep Space Exploration Laboratory, Hefei 230026, China}

\author[0000-0001-5988-2202]{Shiying Lu}
\affil{School of Mathematics and Physics, Anqing Normal University, Anqing 246011, China;
\url{wen@mail.ustc.edu.cn}} 
\affil{Institute of Astronomy and Astrophysics, Anqing Normal University, Anqing 246133, China}
\affil{Key Laboratory of Modern Astronomy and Astrophysics (Nanjing University), Ministry of Education, Nanjing 210093, China}

\author[0000-0002-7660-2273]{Xu Kong}
\affil{Department of Astronomy, University of Science and Technology of China, Hefei 230026, China; \url{xkong@ustc.edu.cn}} 
\affil{School of Astronomy and Space Science, University of Science and Technology of China, Hefei 230026, China}
\affil{Institute of Deep Space Sciences, Deep Space Exploration Laboratory, Hefei 230026, China}

\begin{abstract}
We conduct a systematic robustness analysis of the hybrid machine learning framework \texttt{USmorph}, which integrates unsupervised and supervised learning for galaxy morphological classification. Although \texttt{USmorph} has already been applied to nearly 100,000 $I$-band galaxy images in the COSMOS field ($0.2 < z < 1.2$, $I_{\mathrm{mag}} < 25$), the stability of its core modules has not been quantitatively assessed. Our tests show that the convolutional autoencoder (CAE) achieves the best performance in preserving structural information when adopting an intermediate network depth, $5\times5$ convolutional kernels, and a 40-dimensional latent representation. The adaptive polar coordinate transform (APCT) effectively enhances rotational invariance and improves the robustness of downstream tasks. In the unsupervised stage, a bagging clustering number of $K=50$ provides the optimal trade-off between classification granularity and labeling efficiency. For supervised learning, we employ GoogLeNet, which exhibits stable performance without overfitting. We validate the reliability of the final classifications through two independent tests: (1) the t-distributed stochastic neighbor embedding (t-SNE) visualization reveals clear clustering boundaries in the low-dimensional space; and (2) the morphological classifications are consistent with theoretical expectations of galaxy evolution, with both true and false positives showing unbiased distributions in the parameter space. These results demonstrate the strong robustness of the \texttt{USmorph} algorithm, providing guidance for its future application to the China Space Station Telescope (CSST) mission.
\end{abstract}

\section{Introduction} \label{sec:1}
The morphological characteristics of galaxies are crucial to understanding their formation and evolution. Galaxies exhibit a rich diversity in morphology, including features such as spiral arms, bars, rings, and tidal tails (e.g.,\citealt{Ho+2019,Sofue+2021,ren+2023,Wei+2024}). These large-scale structures provide critical insights into galaxies' interactions, ongoing physical processes, and evolutionary pathways, as reflected by their physical properties like colors, redshifts, masses, and environmental factors (e.g., \citealt{Kauffmann+2004,omand+2014, schawinski+2014, kawinwanichakij+2017,Gu+2018,su+2025}). Consequently, they offer important clues for a comprehensive understanding of galaxy populations and their evolution.

In early citizen science projects such as Galaxy Zoo (GZ; 
\citealt{Lintott+2008, Lintott+2021}), galaxies were classified by visual inspection. With the development of technology, both parametric and non-parametric methods have been introduced for galaxy classification. Parametric modeling employs predefined mathematical functions such as the Sérsic profile \citep{S+1963,Balcells+2003} to parameterize the two-dimensional surface brightness distribution of galaxies, allowing the extraction of structural parameters like the effective radius ($r_\mathrm{e}$), Sérsic index ($n$), and the axis ratio ($b/a$). Parametric methods typically assume that galaxies possess a single, symmetric, and smooth light distribution. In contrast, non-parametric methods avoid assumptions about functional forms and instead compute statistical indicators directly from image pixels to characterize the light distribution. Examples include the CAS system (Concentration $C$, Asymmetry $A$, and Smoothness $S$; \citealt{Conselice+2000, Conselice+2003}), Gini coefficient, $M_{20}$, and $MID$ statistics (e.g., \citealt{Lotz+2004, Lotz+2006, Freeman+2013}). These indicators quantify the complexity and irregularity of galaxy morphologies,  providing objective measures of their structure.

Visual inspection, parametric, and non-parametric methods have proven effective in past large-scale sky surveys.
However, visual inspection is time-consuming and labor-intensive, parametric fitting relies on idealized assumptions, and non-parametric statistics have limited descriptive power. These drawbacks limit their application to large-scale sky surveys in the new era.
Recent development of large-scale sky surveys, including the Sloan Digital Sky Survey (SDSS; \citealt{Stoughton+2002}), 
the Legacy Survey of Space and Time (LSST; \citealt{Abell+2009}), 
the Euclid space telescope (Euclid; \citealt{Euclid+2025}), 
and the forthcoming China Space Station Telescope (CSST; \citealt{CSST+2025}), will generate vast amounts of high-resolution, multi-band galaxy images. The implementation of machine learning (ML) algorithms for automated galaxy image classification therefore becomes paramount.

Convolutional neural networks (CNNs; \citealt{sun+2018}) have shown remarkable ability in the automated classification of galaxy morphologies, as they can effectively capture structural and textural details in images, enabling accurate and efficient identification \citep[e.g.][]{Krizhevsky+2012, Davis+2014, Dieleman+2015, Fernando+2024}. However, as a supervised learning method (SML), the performance of CNNs heavily relies on both the quality and the size of the labeled training dataset. Currently, data annotation remains laborious and costly, mainly dependent on manual labeling. For example, the well-known ImageNet project (\citealt{Deng+2009}) took nearly a decade to complete the annotation of approximately 12 million images. Therefore, unsupervised machine learning (UML) has been widely adopted for the analysis of large-scale unlabeled galaxy datasets since it does not rely on labeled data. UML techniques can directly extract features from galaxy images while simultaneously uncovering intrinsic data structures and patterns, allowing for clustering based on morphological similarity (\citep{Hocking+2018,  Kolesnikov+2023,Tohill+2024}). 
Recently, \citet{Kolesnikov+2023} applied the \texttt{SOMBRERO}  algorithm, an unsupervised clustering method based on EGG morphological features, to classify galaxies in both SDSS and HST datasets.

However, the performance of UML methods heavily depends on the quality of the feature representations. In high-dimensional data domains such as images, feature structures are usually highly complex and typically contain significant noise and redundant information (e.g., \citealt{xue+2023}). Moreover, many UML methods rely on a single clustering strategy, which may yield inconsistent results when different similarity metrics, such as centroid-based, hierarchical, or density-based measures, are employed. These inconsistencies can lead to classification errors and undermine the reliability of the clustering process.

To overcome these limitations, \citet{Song+2024} proposed the \texttt{USmorph} hybrid machine-learning framework, which integrates feature extraction, unsupervised clustering, and supervised classification to reduce reliance on labeled data. The workflow involves: (1) employing a convolutional autoencoder (CAE; \citealt{Massey+2009}) to denoise and compress galaxy images; (2) applying an adaptive polar coordinate transformation (APCT; \citealt{Fang+2023}) to enhance the model’s rotational invariance; and (3) using a bagging-based voting clustering method \citep{Zhou+2022} to group features and generate training labels for a supervised CNN classifier (GoogLeNet; \citealt{szegedy+2015}).
The supervised learning phase employs the GoogLeNet model, which achieved an overall accuracy of approximately 94\% on the validation set.  The classification outcomes align closely with the parametric morphological measurements and non-parametric morphological spatial distributions of galaxies, thereby confirming the classification's reliability. With \texttt{USmorph}, \citeauthor{Song+2024} successfully classified nearly $\sim$100,000 galaxies in the COSMOS field.

 Although \texttt{USmorph} has demonstrated strong performance in applications involving large-scale datasets, its robustness under varying configurations has not yet been systematically quantified. For safe application to future deep- and wide-field surveys such as CSST, it is essential to ensure that the framework can deliver stable and scientifically reliable results. In this work, we conduct a comprehensive robustness analysis of \texttt{USmorph}. We evaluate its sensitivity to the architectural choices of the CAE, the effectiveness of the APCT in preserving morphological information under rotation, the optimization of the number of clusters $K$, and the stability of supervised classification under varying conditions. We also validate the scientific reliability of the final morphological classifications by examining the consistency between the predicted labels and galaxy evolutionary trends in physical parameter space. These parameters include the Sérsic index ($n$), effective radius ($r_e$), Gini coefficient ($G$), and the second-order moment of the brightest 20\% of the flux ($M_{20}$). Through this analysis, we aim to prove the reliability of \texttt{USmorph} and provide a solid scientific basis for its application in future wide-field surveys.

The structure of this paper is as follows. Section~\ref{sample} describes the COSMOS program and the sample selection. Section~\ref{UML} presents the robustness tests of the unsupervised stage. Section~\ref{SML} outlines the configuration of the GoogLeNet framework for supervised classification. Section~\ref{discussion} presents the classification results and discussion. Section~\ref{sec:6} summarizes the main conclusions and gives an outlook for future work. Throughout this paper, we use the AB magnitude system \citep{Oke+1983} and assume a \cite{Chabrier+2003} initial mass function and a standard flat $\Lambda$CDM cosmology with parameters $H_0 = 70$ km s$^{-1}$ Mpc$^{-1}$, $\Omega_m = 0.3$, and $\Omega_\Lambda = 0.7$.

\begin{figure}[htbp]
    \includegraphics[width=0.45\textwidth]{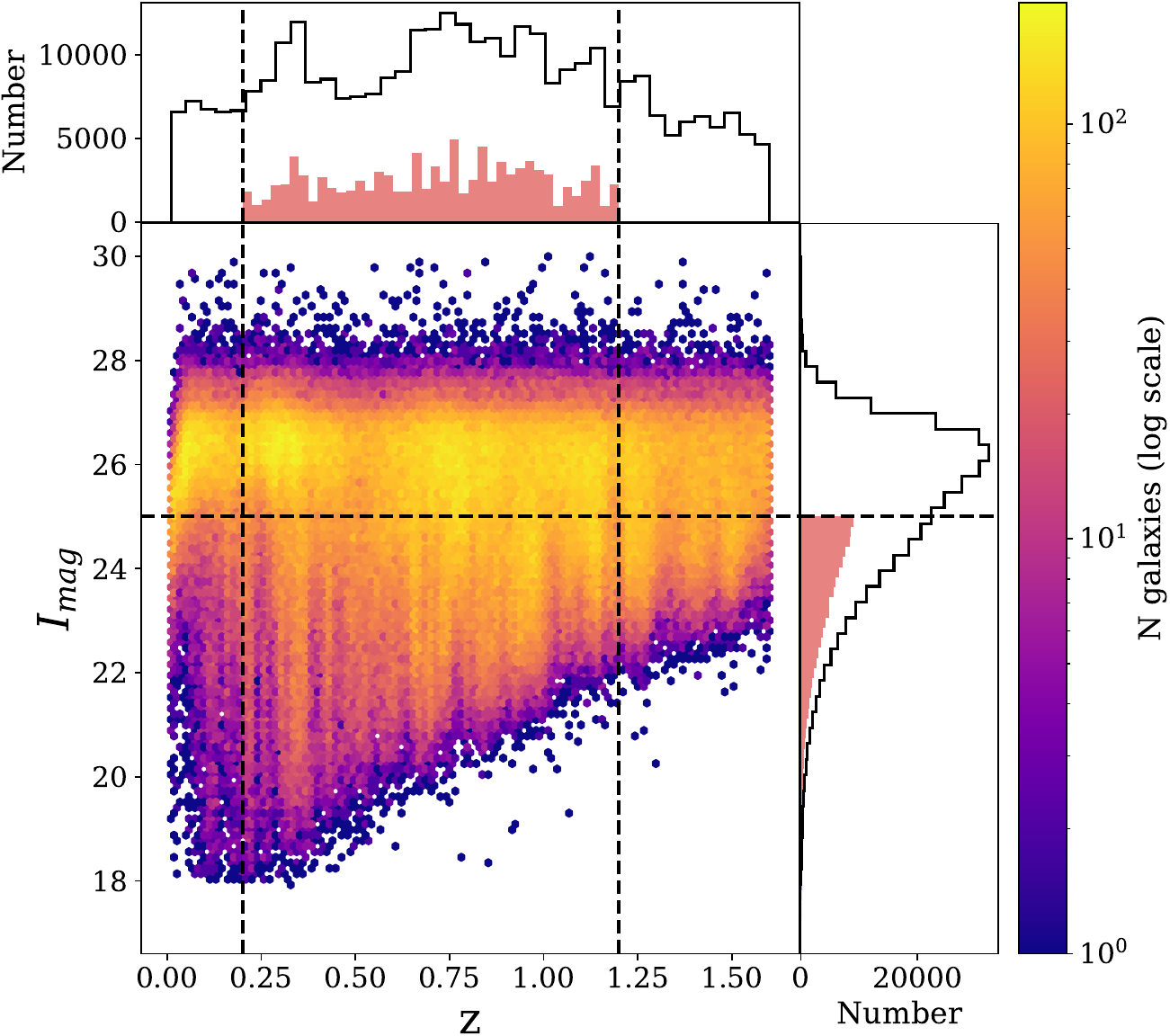}
\caption{The distribution in the $I_{\rm{mag}}$--redshift ($z$) parameter space of galaxies in the COSMOS field. 
Marginalized number distributions along the $I_{\rm{mag}}$ and $z$ dimensions are projected at the top margin and right margin, respectively.  
The subsample satisfying the selection criteria $0.2 < z < 1.2$ and $I_{\rm{mag}} < 25$ is highlighted in red.}      
\label{fig:1}
\end{figure}

\begin{figure*}[htbp]
   \centering
    \includegraphics[width=0.85\textwidth]{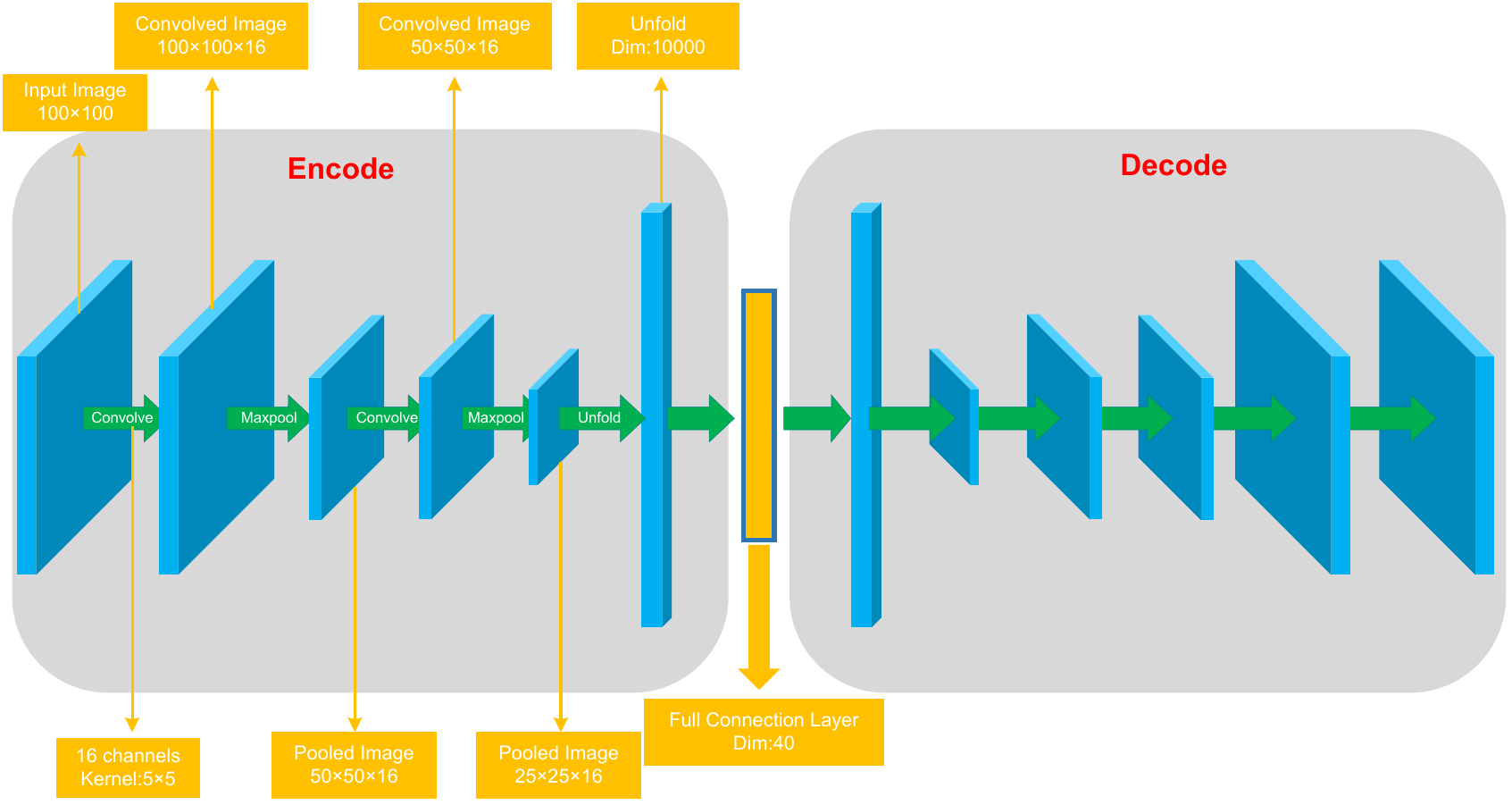}
    \caption{CAE architecture with symmetric encoder-decoder structure. Encoder: $100\times100$ input undergoes two $5\times5$ convolutional layers (16 channels throughout) interleaved with max pooling, producing $25\times25\times16$ features. The resulting feature maps are flattened to a 40-dimensional vector that encodes galaxy features.}
    \label{fig:2}
\end{figure*}

\section{Observation and Sample Seletction} \label{sample}
The COSMOS2020 ``Farmer'' catalog \citep{Weaver+2022} is one of the most comprehensive multiwavelength photometric datasets available for extragalactic studies. It is based on the COSMOS survey \citep{Scoville+2007}, which targets the interplay between galaxy evolution, star formation, active galactic nuclei (AGNs), dark matter, and large-scale structure across a redshift range of $0.5 < z < 6$, over an area of approximately 2 deg$^2$. The dataset covers 35 photometric bands from the ultraviolet to the mid-infrared, incorporating images from both space- and ground-based observatories. In particular, high-resolution imaging from the Hubble Space Telescope (HST) with the Advanced Camera for Surveys (ACS) in the F814W filter covers a contiguous region of $\sim$1.64 deg$^2$, comprising 590 pointings with an average exposure time of 2028 seconds per tile \citep{Koekemoer+2007}. These data were processed using the \texttt{MultiDrizzle} package \citep{Koekemoer+2003}, resulting in final mosaic images with a pixel scale of $0\farcs03$ and a 5$\sigma$ depth of 27.2 AB magnitudes for point sources within a $0\farcs24$ aperture.

In addition to accurate photometric measurements, the COSMOS2020 catalog also provides estimates of various physical properties derived from spectral energy distribution (SED) fitting, such as photometric redshifts, stellar masses, and star formation rates. Two independent SED-fitting codes, {\tt EAZY} \citep{Bramme+2008} and {\tt LePhare} \citep{Ilbert+2006}, were employed to estimate photometric redshifts. In this work, we adopt the redshifts derived from {\tt LePhare}, as they demonstrate better reliability within our magnitude range of interest \citep[see Figure~15 of][]{Weaver+2022}. The redshift estimation process utilizes a library of 33 galaxy templates constructed from the stellar population synthesis models of \citet{Bruzual+2003} and \citet{Ilbert+2009}, along with a variety of dust attenuation prescriptions, including the starburst curve from \citet{Calzetti+2000}, the SMC extinction curve \citep{Prevot+1984}, and two modified versions of the \citet{Calzetti+2000} law incorporating the 2175~\AA{} bump. The final redshift ($z_{\text{LePh}}$) is defined as the median of the redshift likelihood distribution.

\begin{figure*}
   \centering
    \includegraphics[width=0.75\textwidth]{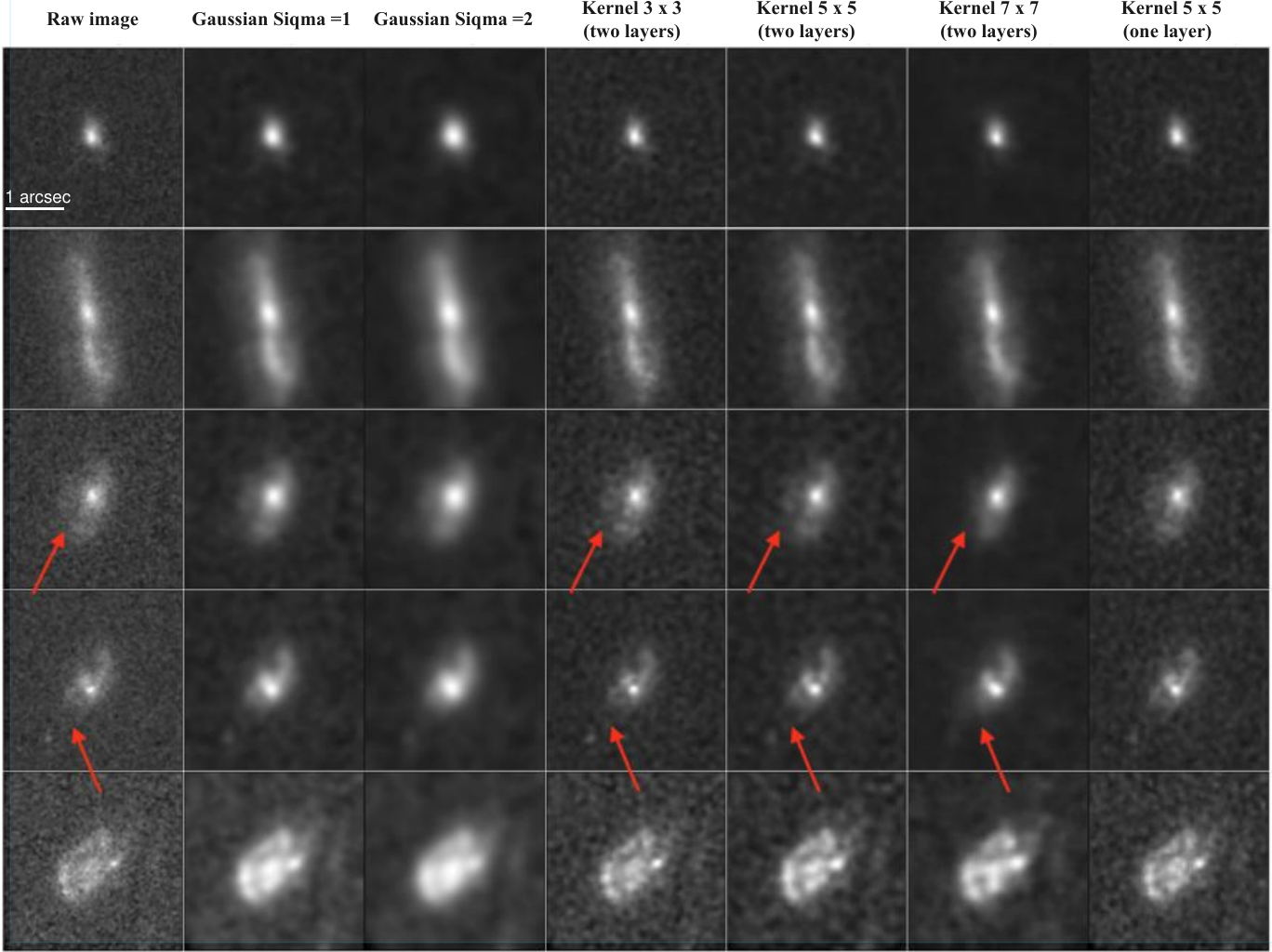}
    \caption{Denoising results using filters of different sizes. From left to right, the images show: the original noisy images; images denoised using Gaussian filtering ($\sigma = 1$ and $2$); images denoised with CAE filters of sizes $3\times 3$, $5\times 5$, and $7\times 7$; and images denoised using only a single convolutional layer. The denoising performance of the $5\times 5$ and $7\times 7$ filter sizes is quite similar; however, some small and faint structures in galaxies may be overlooked when using the $7\times 7$ filter, as indicated by the arrows. \textbf{The white bar in the first panel indicates an angular scale of $1\arcsec$ ($\approx 33$ pixels).}}
    \label{fig:3}
\end{figure*} 

We select the parent sample of galaxies from the COSMOS2020 catalog based on the following criteria:
\begin{enumerate}[label=(\arabic*)]
\item $\rm lp_{type}=0$, to ensure that we are selecting galaxies rather than stars \footnote{For detailed parameter descriptions, please refer to: \href{https://irsa.ipac.caltech.edu/data/COSMOS/tables/cosmos2020/}{COSMOS 2020 ``Farmer’’ Catalog.}};
\item $\rm FLAG_{COMBINE}=0$, to ensure that the flux measurements are not affected by bright stars and that the target objects are not located at the edge of the image, thereby ensuring the reliability of redshift and stellar mass measurements;

\item $0.2 < z < 1.2$, to ensure that galaxy morphology is measured in the rest-frame optical wavelength range of approximately 3700–6800~\AA.

\item $I_{\mathrm{mag}}<25$, to exclude galaxies that are too faint for reliable morphological measurements.

\item signal-to-noise ratio (S/N) greater than 5 and no abnormal pixels, to ensure the high-quality of source images. 
\end{enumerate}

99,806 galaxies were retained in the parent sample. Figure~\ref{fig:1} shows the distribution of these galaxies in the $I_{\rm{mag}}$--redshift ($z$) parameter space.

\section{Analysis for The Unsupervised Method for Morphological Classification} \label{UML}
This section introduces the workflow design and robustness validation of the UML module within the  \texttt{USmorph} framework. The process of the UML module consists of three key stages: (1) using a CAE to reduce the dimensionality of raw galaxy images and extract key morphological features from high-dimensional image data; (2) utilizing APCT to resolve the rotational invariance problem in galaxy classification; (3) employing a bagging-based multi-clustering approach to cluster the galaxy images processed through the aforementioned steps.

\subsection{data pre-processing}
During the data pre-processing stage, we uniformly crop all images to $100 \times 100$ pixels to retain sufficient structural information. Since the cropped size is related to the effective radii of galaxies in our sample, we use the \texttt{GALAPAGOS} software (\citealt{Barden+2012,haubler+2022}) to measure the effective radius of each galaxy. Subsequent statistical analysis shows that 98\% of the sample galaxies have effective radii smaller than 50 pixels, confirming that the adopted $100 \times 100$ pixel cropping size is sufficient for subsequent morphological classification tasks.

\sloppy 
We further apply a max–min normalization pretreatment to each cutout. The flux of each pixel in float type is converted to nonnegative integers by the following procedure:
\begin{equation}
    f_{new}= \left\lfloor \frac{f-f_{min}}{f_{max}-f_{min}} \right\rfloor \times N,
\end{equation}
where $f_{max}$ and $f_{min}$ are the maximum and minimum fluxes of pixels, respectively. 
The floor function, $\lfloor x \rfloor$, gives the greatest integer less than or equal to $x$. 
$N$ is a given integer describing the degree of discretization, which is set as 500 in this paper.
\fussy Our previous work (\citealt{Zhou+2022}) had shown that the choice of N does not have a significant influence on our results. 

\subsection{Convolutional Autoencoder Framework for Noise Reduction}
Convolutional autoencoders (CAE; \citealt{Masci+2011}) represent an extension and refinement of traditional autoencoders (AE; \citealt{Bourlard+1988}). Unlike AEs that rely on fully connected layers for feature extraction, CAEs employ convolutional and deconvolutional layers that enable efficient learning of local structural features within images while preserving spatial information. Therefore, CAEs are widely used in image denoising, data compression, and semantic feature extraction (\citealt{Krizhevsky+2012,Zhou+2022,Dai+2023}).
A CAE model consists of an encoder and a decoder. The encoder utilizes CNN to extract hierarchical features from the input data and progressively compresses the spatial information through a series of convolutional and max-pooling operations. The decoder reconstructs the denoised image from the compressed latent representation via transposed convolutional operations.

\begin{figure*}
   \centering
    \includegraphics[width=0.60\textwidth]{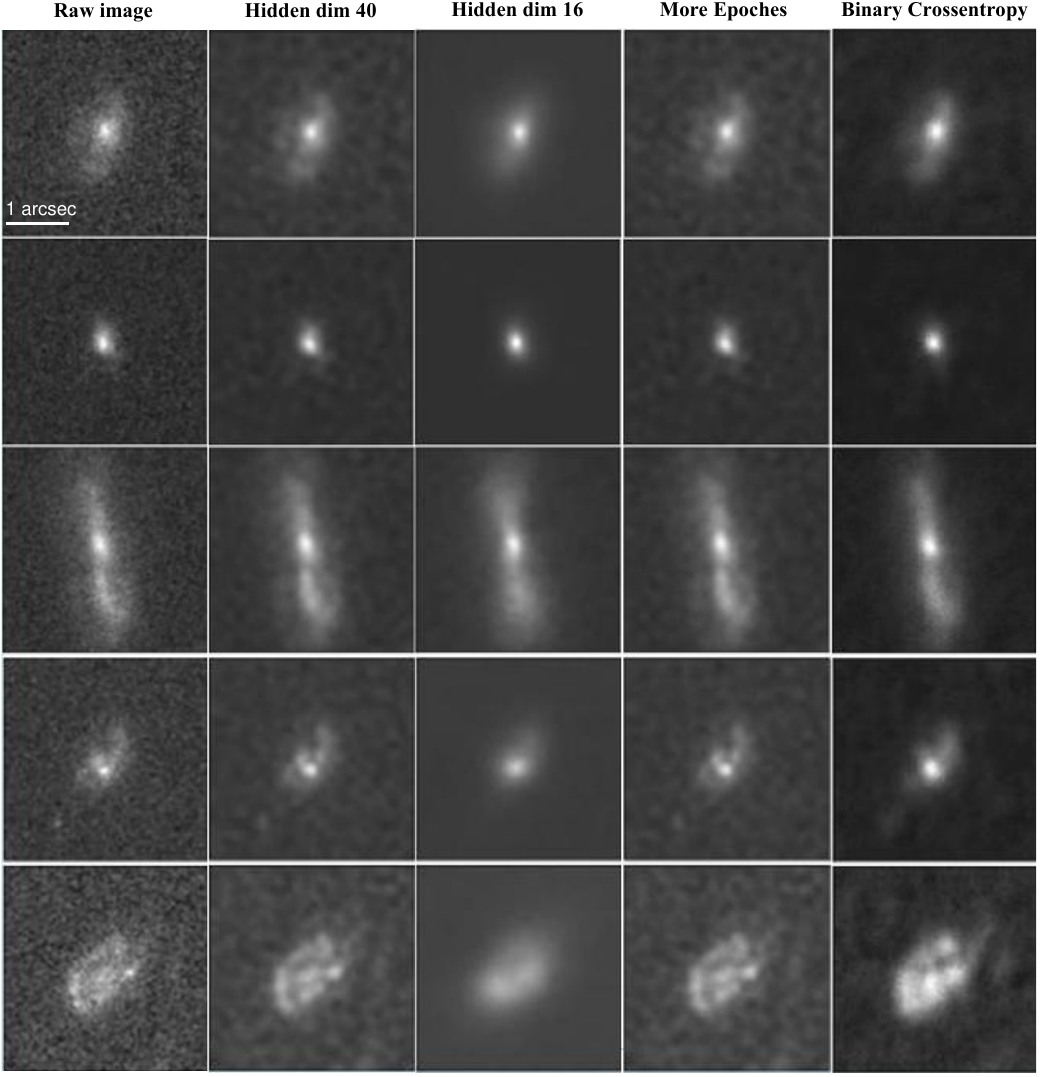}
    \caption{Denoising results obtained with different configurations.  
    The first column shows raw images of five randomly selected galaxies; the second column shows the denoising results with the configuration used in this manuscript ({\tt Hidden dim}=40); the third column shows the denoising results when the hidden dimension is reduced to 16; the fourth column shows the denoising results when 15 more training epochs are added; the fifth column shows the denoising results when using the binary cross-entropy loss function.}
    \label{fig:4}
\end{figure*}

\begin{figure*}
   \centering
   \includegraphics[width=0.85\textwidth]{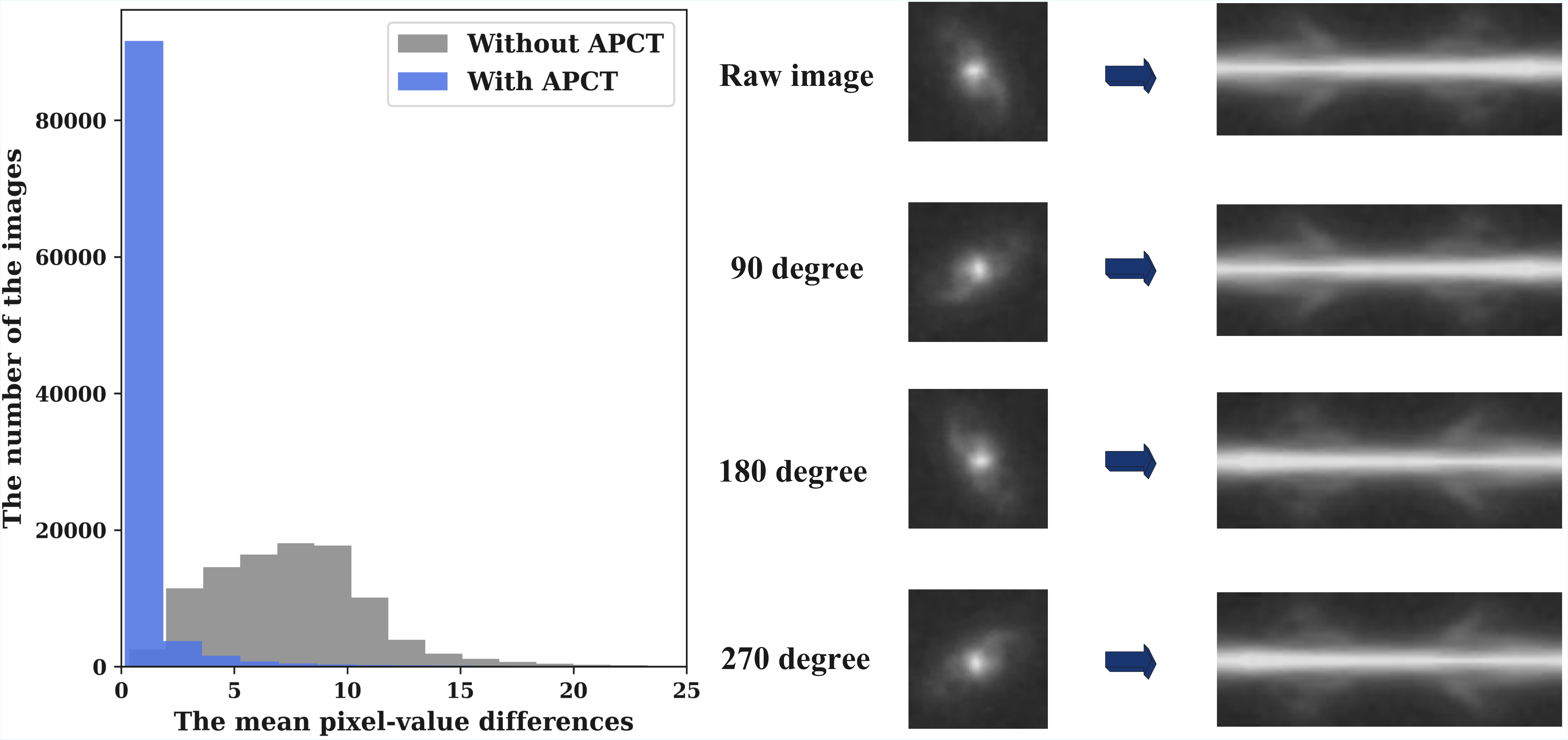}
    \caption{
    \textbf{Left:} Distribution of mean pixel-value differences for galaxy images before and after $90^\circ$ rotation, comparing cases with (blue) and without (grey) APCT. Each value represents the per-galaxy average of absolute differences $|\Delta I(x,y)|$ across all pixel positions. APCT results in near-zero differences, demonstrating strong rotational invariance. In contrast, non-APCT images exhibit significant deviations due to imperfect centrosymmetry. 
    \textbf{Right:} Example galaxy processed with APCT under rotations of $0^\circ$, $90^\circ$, $180^\circ$, and $270^\circ$, showing that the APCT-processed images remain nearly unchanged, regardless of the rotation of the original input.}
   \label{fig:5}
\end{figure*}

In this work, the input data dimension is set to $100 \times 100$ (i.e., 10,000 pixels). The encoder comprises two layers, each containing a convolutional operation followed by a max-pooling step. The output is then flattened and compressed into a 40-dimensional vector via a fully connected layer in order to balance representational capacity and computational efficiency. The overall architecture of the proposed CAE is illustrated in Figure~\ref{fig:2}. 
The model is trained to minimize the reconstruction error between the input and the output images, quantified by a loss function, and to optimize the network parameters using the Adam optimizer \citep{Kingma+2014}. We employ the Rectified Linear Unit (ReLU; \citealt{Agarap+2018}) activation function and the Mean Squared Error (MSE; \citealt{lehmann+2006}) loss, defined as:

\begin{equation}
    \text{loss} = \frac{1}{10000n} \sum_{i=1}^{n} \sum_{j,k}^{100} (\hat{y}_{j,k} - y_{j,k})^2,
\end{equation}
The loss is the mean squared error between all input and reconstructed pixels across all $n$ subsamples in a batch, where $y_{jk}$ and $\hat{y}_{jk}$ are the input and reconstructed pixel values at position $(j,k)$.
In the training phase, a \texttt{batch size} of 8 is used, and the initial learning rate for the CAE is set to $3 \times 10^{-4}$. The learning rate is adjusted dynamically using an exponential decay strategy (\citealt{li+2019}) to facilitate better optimization. The base number of training epochs is set to 32. If the validation loss plateaus, an additional 10 epochs are introduced to further enhance denoising performance, resulting in a total of  42 epochs.

We further conduct comparative experiments to validate the chosen CAE configuration.
Figure~\ref{fig:3} contrasts Gaussian (\citealt{Gonzalez+2006}) blurring with CAE-based denoising. The first column shows the raw images of five randomly selected galaxies; the second and third columns present Gaussian filtering results with standard deviations of $\sigma = 2$ and $\sigma = 3$, respectively; the fourth to sixth columns display denoising outcomes using CAEs with kernel sizes of $3 \times 3$, $5 \times 5$, and $7 \times 7$. As shown in Figure~\ref{fig:3}, Gaussian blurring tends to smear structural features of the galaxies, whereas the CAE approach preserves these features more effectively. For instance, Gaussian filtering significantly increases the apparent size of the first galaxy, while certain features in the fourth and fifth galaxies are excessively smoothed. Kernel size considerably influences denoising performance: smaller kernels (e.g., $3 \times 3$) retain more residual noise, while larger kernels (e.g., $7 \times 7$) suppress faint structures, as indicated by the red arrows. The seventh column presents results from a CAE using a single $5 \times 5$ convolutional layer, which performs comparably to a two-layer CAE with $3 \times 3$ kernels. This suggests that increasing network depth can be functionally similar to enlarging the receptive field via larger kernels. However, we settle on an encoder comprising two convolutional layers with $5 \times 5$ kernels since very deep networks may also overlook faint features.

When selecting the output dimensionality of the CAE, a trade-off must be made between computational efficiency and the preservation of image features: excessively high dimensionality increases the computational cost of downstream clustering tasks, while overly low dimensionality may result in the loss of key morphological features of galaxies. \citet{Zhou+2022} found that increasing the dimensionality to 100 preserves all features but only provides limited improvement in classification performance. Therefore, we adopt a 40-dimensional representation in this work, which maximizes computational efficiency while retaining the essential morphological information of galaxy images. In Figure~\ref{fig:4}, the first column shows the raw images of 5 randomly selected galaxies; the second column presents the denoised results obtained with the adopted CAE configuration (hidden dim=40); and the third column shows the reconstruction when the latent dimensionality is set to 16. A comparison clearly demonstrates that an excessively small dimensionality leads to substantial feature loss. 

We further conduct several other comparative experiments: the fourth column displays the denoised results after training for an additional 25 epochs beyond loss stabilization (i.e., a total of 15 extra epochs), which show no evident improvement over the second column; the fifth column shows the results obtained when an alternative loss function is used (binary cross-entropy), which fails to visually outperform the adopted configuration because it loses certain galaxy details. Figures~\ref{fig:3} and \ref{fig:4} further validate the robustness of our parameter choices.

\subsection{The APCT  Preprocessing Strategies} 
Convolutional Neural Networks (CNNs) possess translation invariance but lack rotational invariance in image classification. This deficiency can cause misclassifications of rotated galaxy images that share identical morphology \citep{yao+2019}. 
Traditional approaches, such as ``rotated image augmentation,'' enhance rotational robustness by substantially expanding the training set (e.g., adding $2$--$8$ rotated versions per image; \citet{Dominguez+2018}). However, this method lacks efficiency and relies on similarity between the angular distributions of the training and test sets. It fails to handle the scenario where galaxy orientations are randomly distributed in real observations.

To ensure rotationally invariant classification of galaxy morphology, we employ Adaptive Polar Coordinate Transformation (APCT; \citet{Fang+2023}). Conventional polar coordinate transformations utilize a fixed polar axis, which can induce horizontal shifts in the transformed images when rotating the original images. In contrast, APCT defines the coordinate system using a rotationally invariant polar axis instead of a fixed one. The technique first identifies the brightest and darkest pixels within the image to determine the direction of the polar axis. The axis is then rotated counterclockwise in  \(0.05\) rad steps. Subsequently, all pixels are mapped (``stacked'') onto the polar coordinate space along the rotated axis. Finally, a mirroring operation is applied to the image to accentuate central features. 
This process enhances the model's rotational invariance and simultaneously improves the saliency of morphological features within the image.

Figure~\ref{fig:5} illustrates the distribution of the mean pixel-value differences before and after a 90$^\circ$ rotation of galaxy images, with and without APCT. The results show that, with APCT, the majority of images exhibit minimal pixel-value differences, whereas without APCT, the differences are significantly larger due to the lack of perfect central symmetry in galaxies. This result demonstrates that the APCT effectively eliminates the effect of rotation on pixel distribution while preserving the core features of the image by converting Cartesian coordinates to polar coordinates. By transforming the problem of rotation invariance into one of translation invariance, the APCT can significantly enhance the rotational classification robustness of CNNs, which will be further validated in ~\ref{sec:5.1} via more tests.

\subsection{Voting strategy and Post-hoc label alignment}
Hybrid clustering voting boosts (\citealt{breiman+1996}) galaxy classification robustness and accuracy by integrating multiple models. Given that galaxies frequently exhibit transitional morphologies during their evolution, and different algorithms employ distinct similarity criteria, individual methods struggle with ambiguous samples from evolutionary sequences. Our Bagging-based multi-model voting approach minimizes errors by excluding samples with significant inter-model disagreement (\citealt{Zhou+2022}).

The specific implementation combines three clustering algorithms: \textit{K}-means Clustering (K-means; \citealt{Hartigan+1979}), Hierarchical Balanced Iterative Reduction and Clustering Algorithm (BIRCH; \citealt{Zhang+1996}), and the Agglomerative Clustering algorithm (Agg; \citealt{Murtagh+1983, Murtagh+2014}). This collective approach reduces the inherent errors of individual models. Samples achieving unanimous agreement among all three algorithms constitute the final consensus set, while discordant samples are excluded. The schematic diagram is shown in Figure 4 of \citet{Zhou+2022}.

We choose these three clustering algorithms due to their complementary properties: K-means is particularly effective for convex datasets; Agg captures hierarchical structures and performs well on concave distributions; and Birch efficiently handles noisy and large-scale data. Specifically, K-means rapidly partitions data into initial clusters based on distance metrics; Agg identifies hierarchical relationships within the data; and Birch processes large datasets efficiently by constructing a clustering feature tree.

By integrating the clustering results of all three algorithms through a voting mechanism, we effectively minimize the bias introduced by individual algorithms in the galaxy morphology classification task, substantially improving clustering purity (\citealt{Zhou+2022,Dai+2023,liu+2023}).
In contrast, other algorithms, such as density-based spatial clustering of applications with noise (DBSCAN;\citealt{Deng+2020}) and Mean shift clustering (Mean-shift; \citealt{Comaniciu+2002}), present challenges for application in astronomical survey projects with massive data due to their parameter sensitivity, cumbersome parameter adjustment, and insufficient efficiency in processing large-scale data.

After determining the voting strategy, we adopt a post-hoc label alignment approach (\citealt{liu+2023}), which consists of two steps: over-clustering and visual category mapping. In unsupervised image classification tasks, clustering partitions samples solely based on feature similarity, and its inherent "semantic deficiency" makes the results difficult to directly correspond to physical research objectives. Therefore, the reasonable selection of the number of clusters is crucial. A common practice is to preset a relatively large number of categories based on scientific goals, allowing the unsupervised algorithm to fully capture fine-grained feature differences, and then assign semantic information to the clustering results through a small number of visual annotations.

In this study, we set $k = 50$ based on the following considerations. Firstly, $k=50$ is sufficient to ensure that each subcategory is highly homogeneous in morphology. 
Our tests demonstrate that a set of 50 clustering samples can typically be annotated within five minutes, without imposing excessive manual effort.
Secondly, a larger number of categories helps to characterize subcategory structures and identify special celestial bodies (such as tidal tails, strong gravitational lenses, and littlered dots), thereby providing deeper support for subsequent scientific analysis\footnote{See Figure 4 of \citet{Song+2024} for Post-hoc label alignment.}.

The 63,202 galaxies receive inconsistent votes with different clustering algorithms, accounting for 64\% of the samples.
This relatively high fraction stems primarily from discrepancies in the voting results among different clustering algorithms. This indicates that these galaxies exhibit ambiguous or transitional morphological features, which preclude consistent classification. 
After excluding samples with inconsistent votes, a total of 36,604 galaxies were successfully classified into 50 distinct groups. 
Ultimately, the remaining 36,604 galaxies were categorized into five morphological classes: spherical (SPH), early-type disk (ETD), late-type disk (LTD), irregular (IRR), and unclassified (UNC). 

\begin{figure*}
   \centering
   \includegraphics[width=1.0\textwidth]{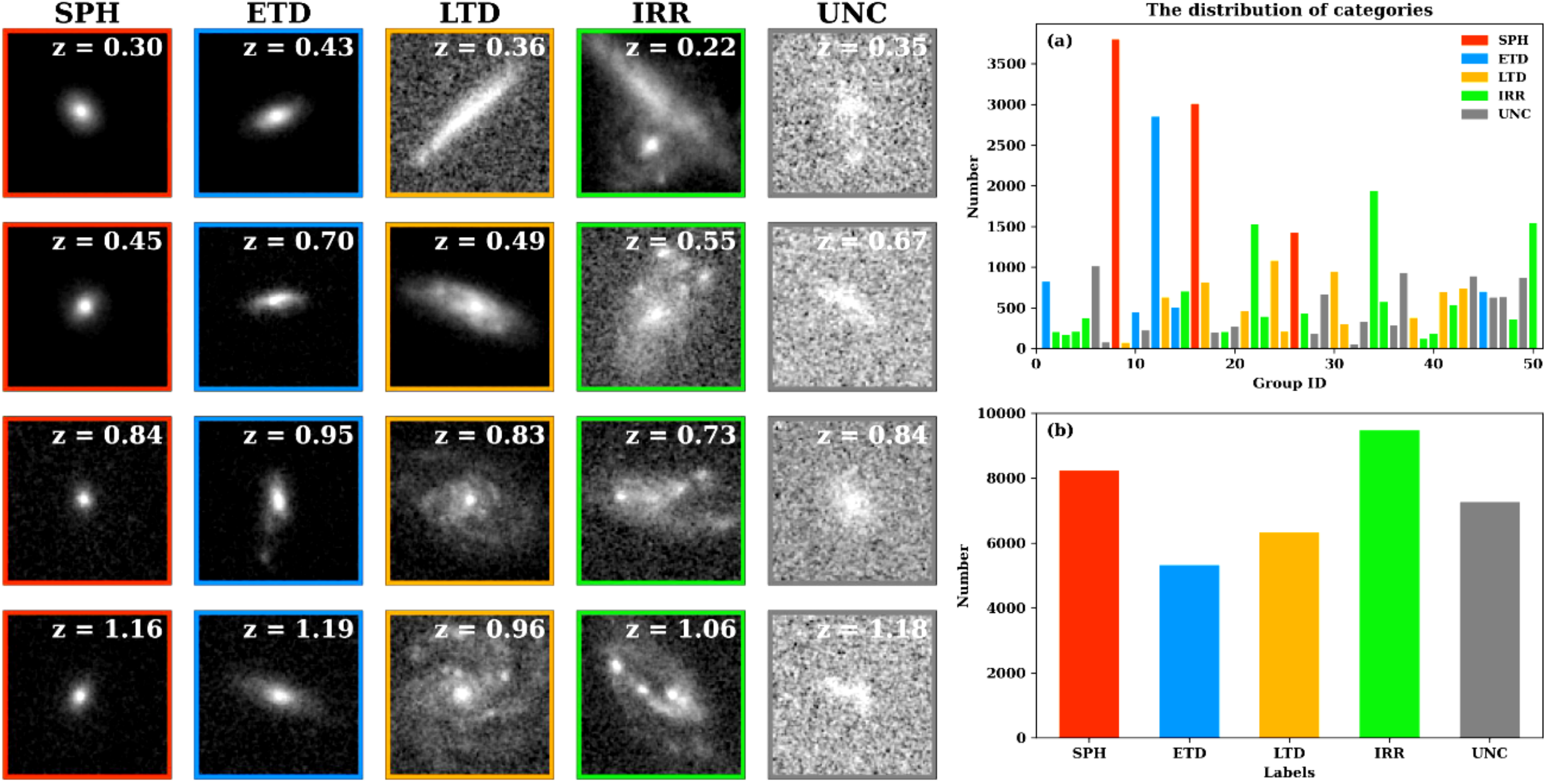}
   \caption{An overview of the UML dataset. \textbf{Left:}: cutouts of galaxies selected from different categories, i.e., SPH, ETD, LTD, IRR, and UNC galaxies, are shown from left to right. \textbf{Right:} the UML dataset illustrates the statistical distribution features of 50 galaxy groups (panel \textbf{(a)}) and the corresponding five-class merger label distributions (panel \textbf{(b)}).
   }
   \label{fig:6}
\end{figure*}

Figure \ref{fig:6} presents the results of UML clustering along with some representative samples. The effectiveness of these 50 clusters will be validated in subsequent experiments. The final sample comprises: 8,233 SPH, 5,322 ETD, 6,320 LTD, 9,468 IRR, and 7,261 UNC galaxies.

\begin{figure}
    \includegraphics[width=0.45\textwidth]{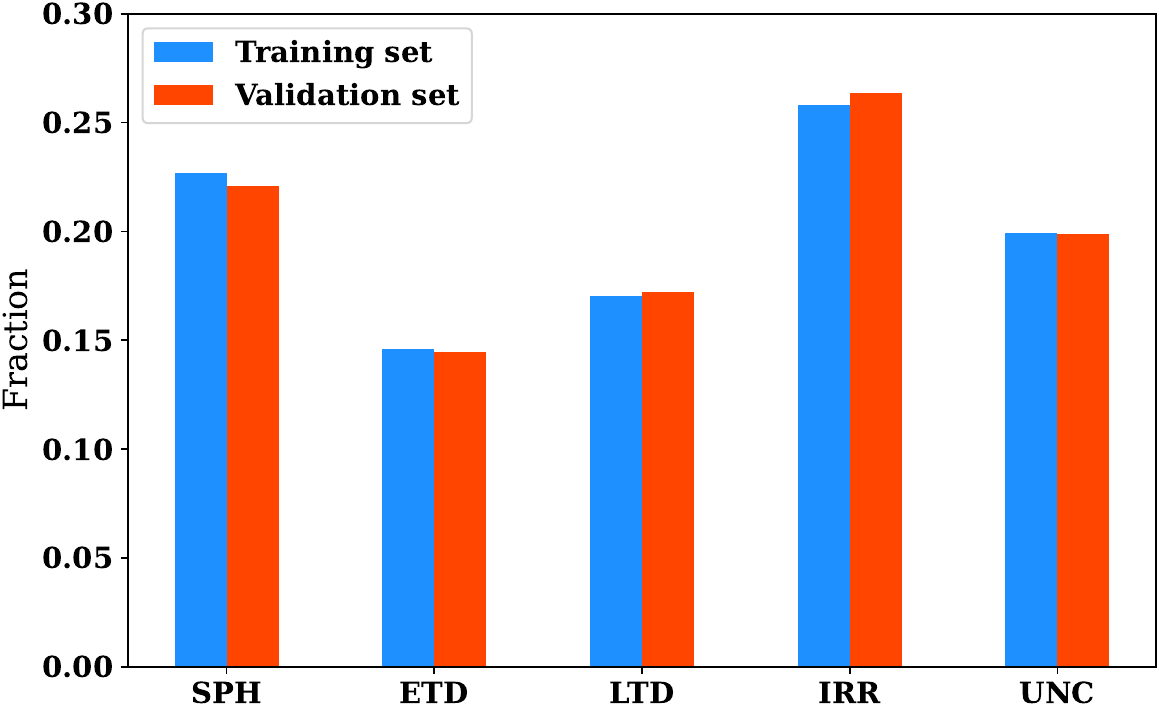}
\caption{The distribution ratio of different types of galaxies in the training and validation sets remains consistent. This consistency ensures the reliability and generalization ability of the model evaluation results.}    
\label{fig:7}
\end{figure}

\begin{figure*}
    \includegraphics[width=2\columnwidth]{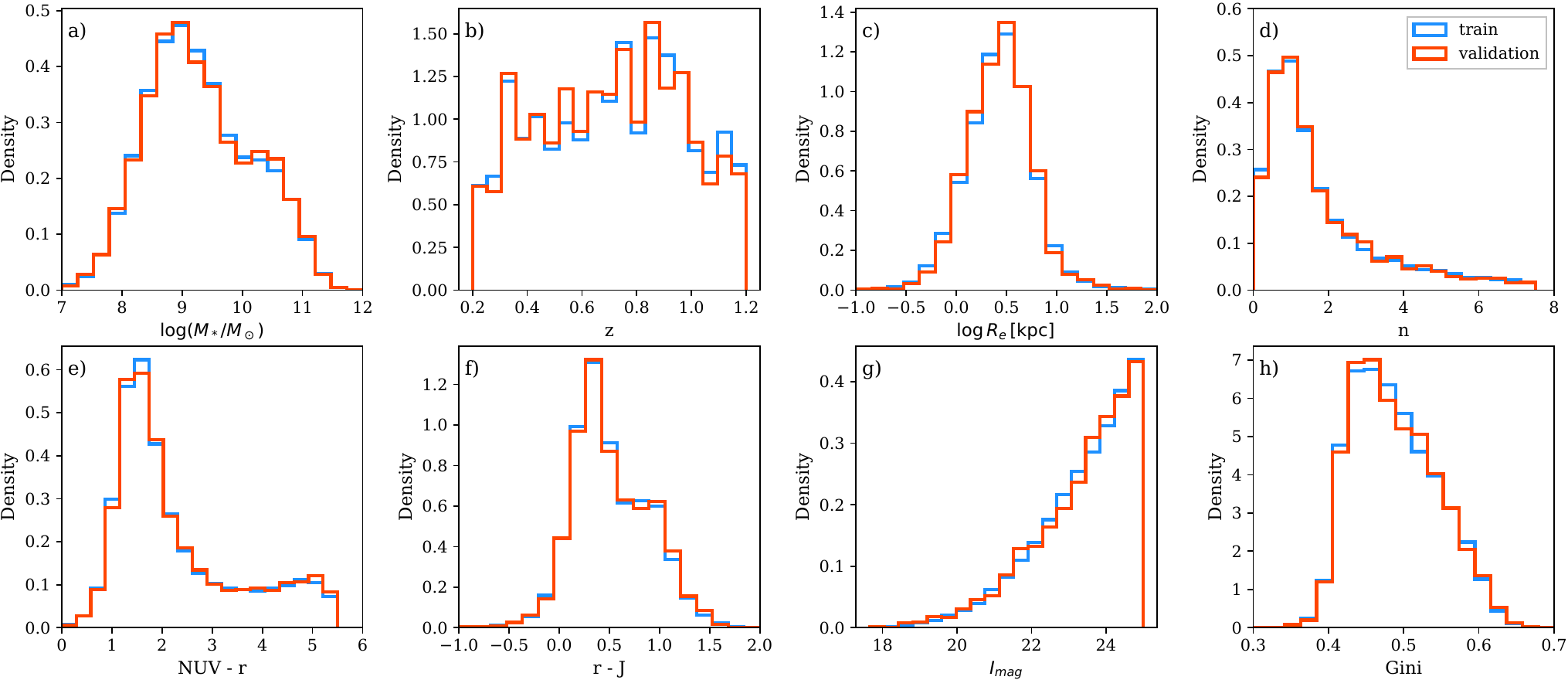}
    \caption{The distribution of training (red) and validation (blue) set data across various physical parameters. The consistent distribution between the training and validation sets illustrates the reliability of our methodology.}
    \label{fig:8}
\end{figure*}

\section{Analysis for The SML Models and the Experiment Settings}\label{SML}
To achieve automated galaxy morphology classification, we utilize a training set of 36,604 reliably labeled sources from UML clustering. Using this dataset, we train and optimize the SML algorithm. This section details the implementation specifics and experimental configuration of the SML algorithm.

\subsection{The GoogLeNet Model}
The GoogLeNet Model (\citealt{szegedy+2015}), a deep convolutional neural network, employs the innovative Inception module, which features a multi-scale convolutional parallel architecture. This design elegantly achieves a balance between network depth, width, and computational efficiency, demonstrating exceptional performance across a wide range of application scenarios. The GoogLeNet architecture uses nine sequential inception modules, each applying parallel $1 \times 1$, $3 \times 3$, and $5 \times 5$ convolutions. This design captures broad image areas while preserving fine details. With increasing network depth and width, GoogLeNet can achieve greater parameter efficiency, a reduction in total parameters, and improvements in both performance and computational efficiency.
In a comparative analysis, \citet{Fang+2023} demonstrated that GoogLeNet exhibited the highest stability among three deep learning frameworks for galaxy image classification.

\subsection{Experiment Setting}
We utilize 36,604 reliably labeled samples obtained from UML clustering as the training set to classify the remaining 63,202 galaxies using SML. 
In principle, the entire UML dataset should be employed for training to enable the algorithm to fully exploit all available information, but this may lead to overfitting.
Taking the large sample size, we divide the UML dataset into training (Tra) and validation (Val) sets at a 9:1 ratio.

We further verify that key feature distributions were balanced between the two sets to ensure the model’s generalization capability and the reliability of performance evaluation, as well as to avoid potential biases caused by discrepancies in data distribution. As illustrated in Figure~\ref{fig:7}, the distributions of galaxy morphological types are consistent between the training and validation sets. Similarly, Figure~\ref{fig:8} shows that their distributions in the parameter space of physical properties—such as stellar mass ($M_\ast$), redshift ($z$), and other morphological parameters are highly similar (see Section \ref{sec:5.3}). 
The Kolmogorov–Smirnov (K–S; \citealt{Kolmogorov+1933}) test results (Table~\ref{tab1}) show that all the examined parameters yield $p$-values greater than 0.05, indicating that the training and validation sets follow statistically consistent distributions.

\begin{table}[ht]
\centering
\caption{Results of Partial K-S Test}\label{tab1}
\centering
\setlength{\tabcolsep}{5pt}
\begin{tabular}{lcccc} 
\hline\hline
parameter & \multicolumn{4}{c}{Parameters} \\ 
\cline{2-5} 
& stellar mass & redshift & $\rm R_e$ & n \\ 
\hline
p-value & 0.82 & 0.08 & 0.34 & 0.64 \\ 
\hline
& NUV-r & r-J & $\rm I_{mag}$ & Gini \\ 
\hline
p-value & 0.34 & 0.33 & 0.73 & 0.64 \\ 
\hline
\end{tabular}
\begin{tablenotes}
    \item[] With $p>0.05$, indicating that the training and validation sets follow the same distribution for the corresponding parameter.
  \end{tablenotes}
\end{table}

\begin{table}[htbp]
  \centering 
  \caption{The confusion matrix\label{tab2}} 
  \begin{tabular}{lccc}
    \toprule
    \toprule
    Galaxy Type & Recall  & Precision & F1-Score \\
    \midrule
    SPH & 98.8\% & 99.4\% & 99.0\% \\
    ETD & 95.4\% & 93.8\% & 94.6\%\\
    LTD & 91.6\% & 91.0\% & 91.3\% \\
    IRR & 93.5\% & 91.0\% & 92.2\% \\
    UNC & 91.8\% & 96.2\% & 94.0\% \\
    \bottomrule
  \end{tabular}
  \begin{tablenotes}
    \item[] Note: The overall accuracy of the GoogLeNet model is above 94\%, which proves the robustness of the galaxy classification task. 
  \end{tablenotes}
\end{table}

\begin{figure*}
    \includegraphics[width=2\columnwidth]{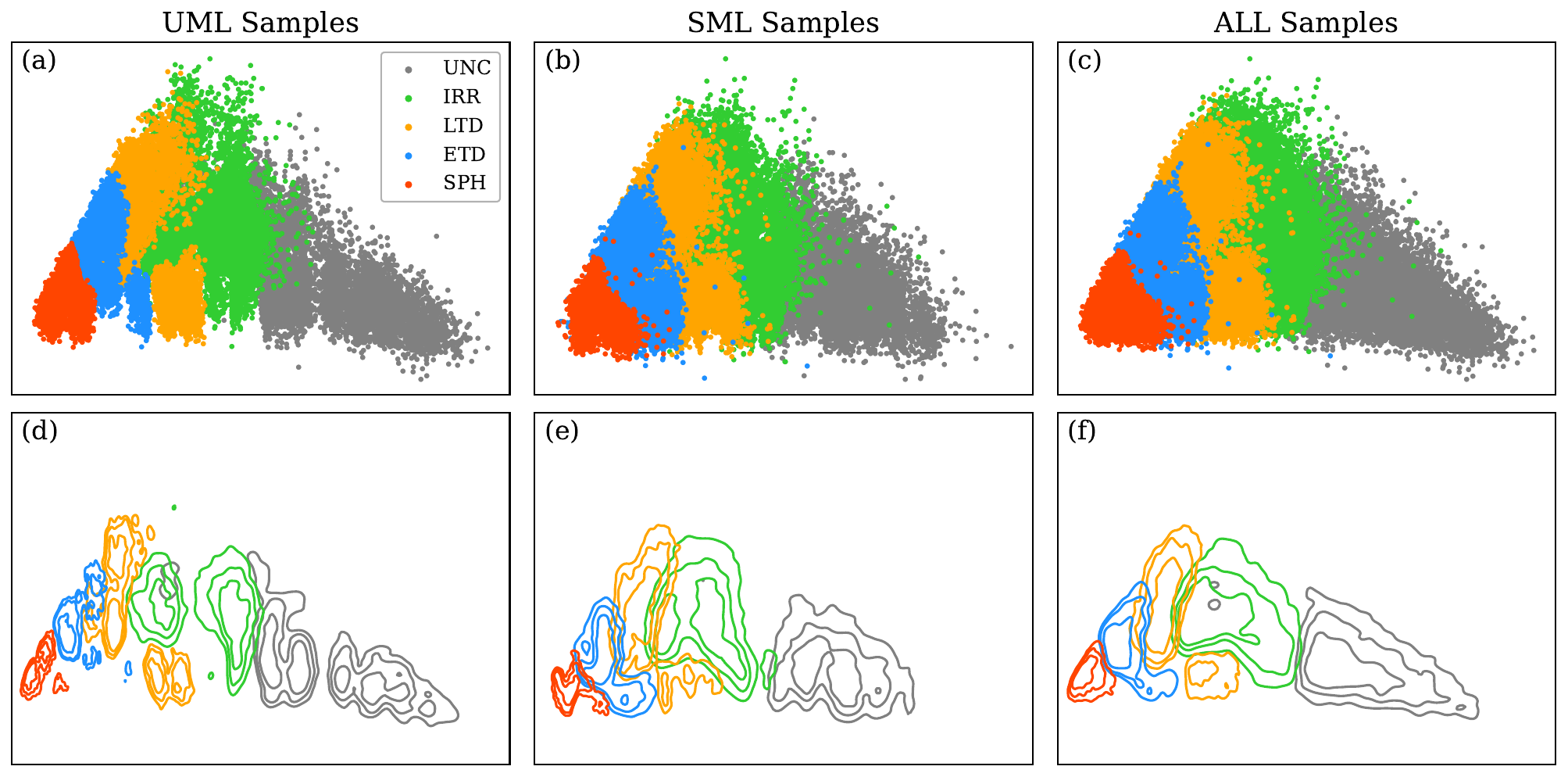}
    \caption{Panels \textbf{(a)}, \textbf{(b)}, and \textbf{(c)} show the t-SNE diagrams of galaxies in the samples obtained using UML, SML, and the combined method, respectively. Panels \textbf{(d)}, \textbf{(e)}, and \textbf{(f)} present the corresponding class-wise contours for each of the three diagrams, enclosing 20\%, 50\%, and 70\% of the respective samples. The clear boundaries between different galaxy types in the t-SNE maps indicate that the classification algorithms (UML, SML, and the combined model) exhibit strong robustness.}
    \label{fig:9}
\end{figure*}

\begin{figure*}
    \includegraphics[width=2\columnwidth]{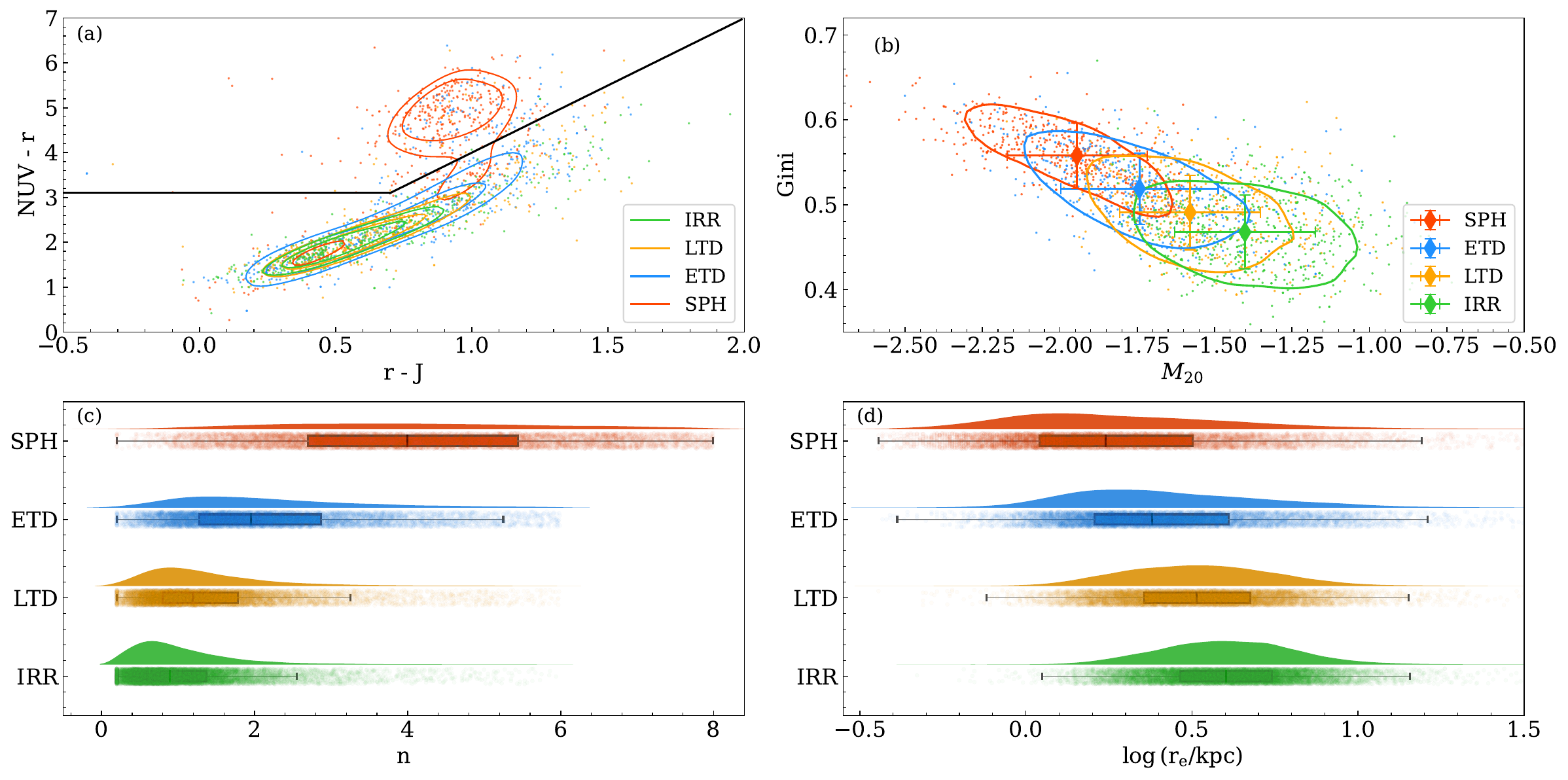}
        \caption{Panels \textbf{(a)} and \textbf{(b)} show the distribution of galaxies in rest-frame $NUV - r$ vs. $r - J$ color space and in the $G - M_{20}$ parameter space, respectively. The contours enclose 70\% of the galaxies in each subclass. The scattered points represent 500 galaxies randomly selected from each subclass. Colored diamonds indicate the median values of the distributions, and the associated error bars represent the standard deviations. Panels \textbf{(c)} and \textbf{(d)} display the distributions of Sérsic index ($n$) and effective radius ($r_{\text{e}}$), respectively. In the box plots, the boxes span the interquartile range (IQR) from the first to the third quartile, the central black line marks the median, and the whiskers extend to the minimum and maximum values.}
    \label{fig:10}
\end{figure*}

We take a \texttt{batch size = 32} during the supervised training phase. The \texttt{learning rate} is reduced to 0.0001, which facilitates stable convergence of the model. To ensure stable and efficient convergence during training, the GoogLeNet model is optimized using stochastic gradient descent (SGD; \citealt{Sutskever+2013}) with momentum.
We use a maximum of 300 training epochs. If the overall accuracy does not show a significant improvement after 150 epochs, the training will be terminated to mitigate overfitting and avoid waste of computational resources. We perform training on a server with an \texttt{NVIDIA A10} GPU. Our configuration achieves a balance between training efficiency and generalization capability within the constraints of available video memory capacity. The overall parameter setup emphasizes the integration of performance efficacy and system stability.

\subsection{Performances of SML Models}
To verify the effectiveness of the GoogLeNet model in galaxy image classification, we employ \textit{Precision}, \textit{Recall}, \textit{F1-Score}, and \textit{Accuracy} as performance evaluation metrics. All the aforementioned metrics are calculated based on the counts of true positives (TP), false positives (FP), true negatives (TN), and false negatives (FN), with the formulas given below:
\begin{equation}
\textit{Precision} = \frac{TP}{TP + FP}
\end{equation}

\begin{equation}
\textit{Recall} = \frac{TP}{TP + FN}
\end{equation}

\begin{equation}
\textit{F1-Score} = 2 \times \frac{\textit{Precision} \times \textit{Recall}}{\textit{Precision} + \textit{}{Recall}}
\end{equation}

\begin{equation}
\textit{Accuracy} = \frac{TP + TN}{TP + FP + TN + FN}
\end{equation}
Table~\ref{tab2} shows GoogLeNet's excellent performance, achieving an overall accuracy of 94\%. To evaluate the model's generalization ability and detect potential overfitting, we allocate half of the validation set as an independent test set. Simultaneously, we assess the stability of model evaluation across different data partitions by repeated random subsampling.   
The accuracy (Acc) on both the validation and test sets is largely unaffected by variations in the split ratio, as shown in Table~\ref{tab3}.
Our tests show that the GoogLeNet model consistently produces stable feature representations. These results collectively provide evidence that no overfitting was observed during training.

\begin{deluxetable}{cccc}
\tablecaption{Model Accuracy Across Different Data Split Ratios\label{tab3}}
\tablehead{
\colhead{Division} & \colhead{(Tra Set) TA} & \colhead{(Val Set) TA} & \colhead{(Test Set) TA}
}
\startdata
$7:1.5:1.5$ & 100\% & 93.8\% & 93.4\% \\
$8:1:1$ & 100\% & 93.9\% & 93.8\% \\
$9:0.5:0.5$ & 100\% & 95.3\% & 94.1\% \\
\enddata
\begin{tablenotes}
    \item[] Note: To obtain an unbiased evaluation, we introduce a test set to ultimately verify the model's true generalization ability and detect overfitting.
  \end{tablenotes}
\end{deluxetable}

\section{RESULTS AND DISCUSSION}\label{discussion}
In this section, we demonstrate the effectiveness of the unsupervised processing strategy, present the complete results of galaxy morphological classification, and verify their effectiveness through a series of tests.

\subsection{The effectiveness of CAE and APCT}\label{sec:5.1}
To quantitatively evaluate the effectiveness of data processing,  we performed Bagging clustering analysis on two types of images: original images and processed images. Based on the reliable labels obtained from the clustering analysis, we computed the overall classification accuracy of the GoogLeNet model on the validation set under test conditions including image rotation angles of $0^\circ$, $90^\circ$, $180^\circ$, and $270^\circ$. The results are summarized in Table~\ref{tab4}.

Table~\ref{tab4} demonstrates that data processing effectively enhances the classification performance of GoogLeNet. When the validation images are not rotated, the performance gap between models trained with and without data processing is relatively small. This is because the angular distribution difference between the validation set (randomly sampled from the UML dataset) and the training set is minimal. In contrast, rotating the validation images introduces a distribution shift in orientation between the training and validation sets. Under such conditions, the accuracy of the model trained without data processing decreases significantly, while the model trained on processed images maintains stable performance. 
These results align with \citet{Fang+2023} and \citet{Zhang+2024}, further validating the effectiveness of the adopted data processing strategy.

\begin{table}[htbp]
\caption{The validation accuracy under different data processing methods and at various rotation angles.}
\centering
\small
\setlength{\tabcolsep}{2.4pt}       
\renewcommand{\arraystretch}{1.05}   
\begin{tabular}{@{} l c r r r r @{}}  
\toprule[0.8pt]\midrule             
Method & dataset Count & 0° & 90° & 180° & 270° \\ 
\midrule
Raw image & 39,896 & 90.5\% & 86.9\% & 88.9\% & 86.5\% \\
Processed image & 36,604 & 94.4\% & 93.9\% & 93.7\% & 93.8\% \\
\bottomrule[0.8pt]
\end{tabular}
\label{tab4}
\end{table}

\begin{figure*}
    \centering 
    \includegraphics[width=1.0\textwidth]{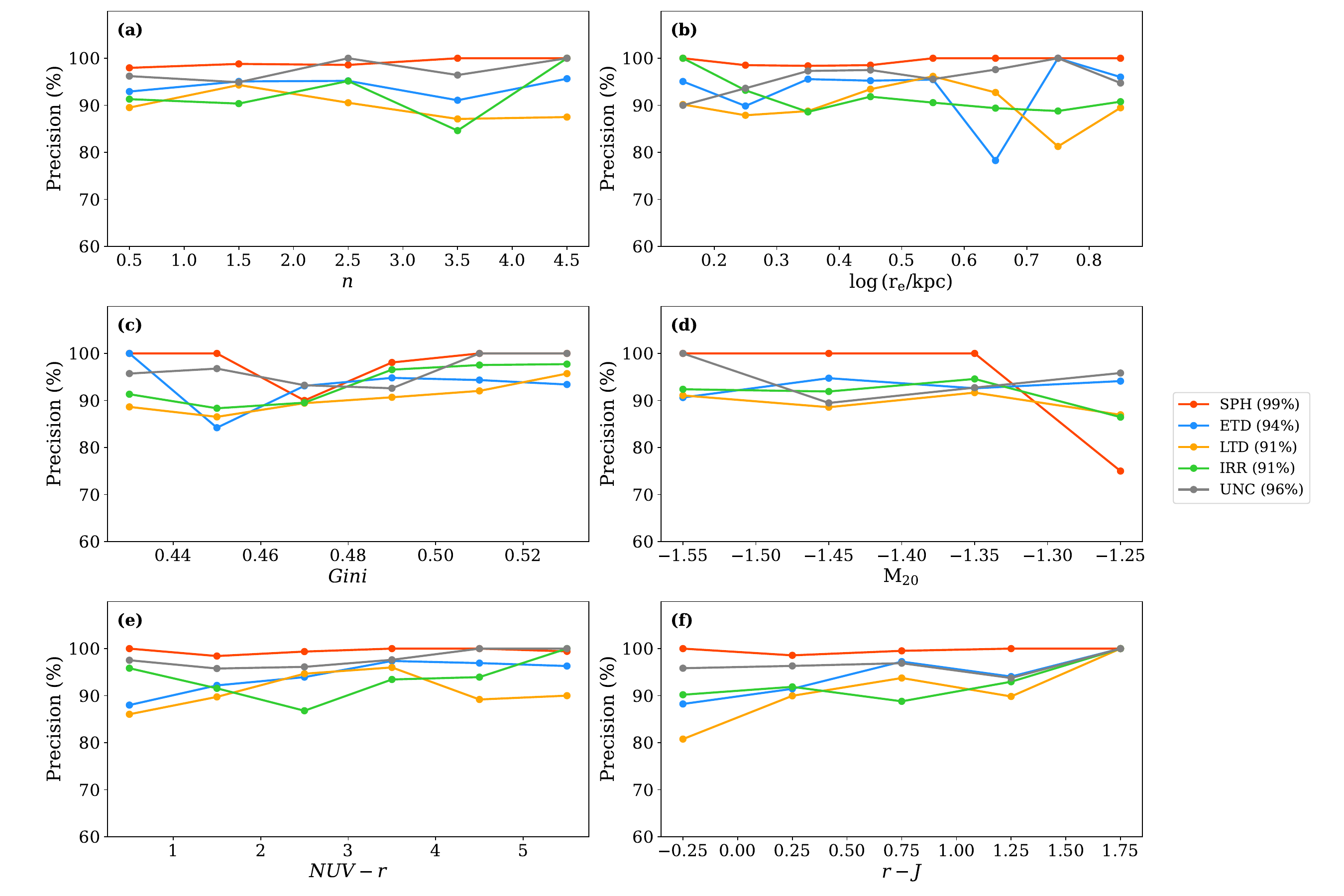}
    \caption{The precision of each category of galaxies in the parameter space of the validation set (red: SPH, blue: ETD, yellow: LTD, green: IRR, and grey: UNC). Panels \textbf{(a)}, \textbf{(b)}, \textbf{(c)}, \textbf{(d)}, \textbf{(e)}, and \textbf{(f)} represent the precision of galaxies on parameters such as Sersic index ($n$), radius ($r_e$), Gini($G$), M\textsubscript{20}, and $NUV-r$ and $r-J$ colors. 
    We define the parameter range to encompass all galaxy types and calculate the median precision of each parameter interval within a metric for its classification precision.
    The parameter range includes all galaxy types, and the median precision within each parameter interval is calculated as a metric for its classification precision.
    The label gives the median precision of this type of galaxy within the entire parameter range. The precision of the model for classifying various categories of galaxies is independent of galaxy morphology and physical parameters. We find no significant difference or trend in the precision of each category in the parameter space, indicating that the False Positives do not rely on galaxy morphological and physical parameters.}
    \label{fig:11}
\end{figure*}

\begin{figure*}
    \includegraphics[width=1.0\textwidth]{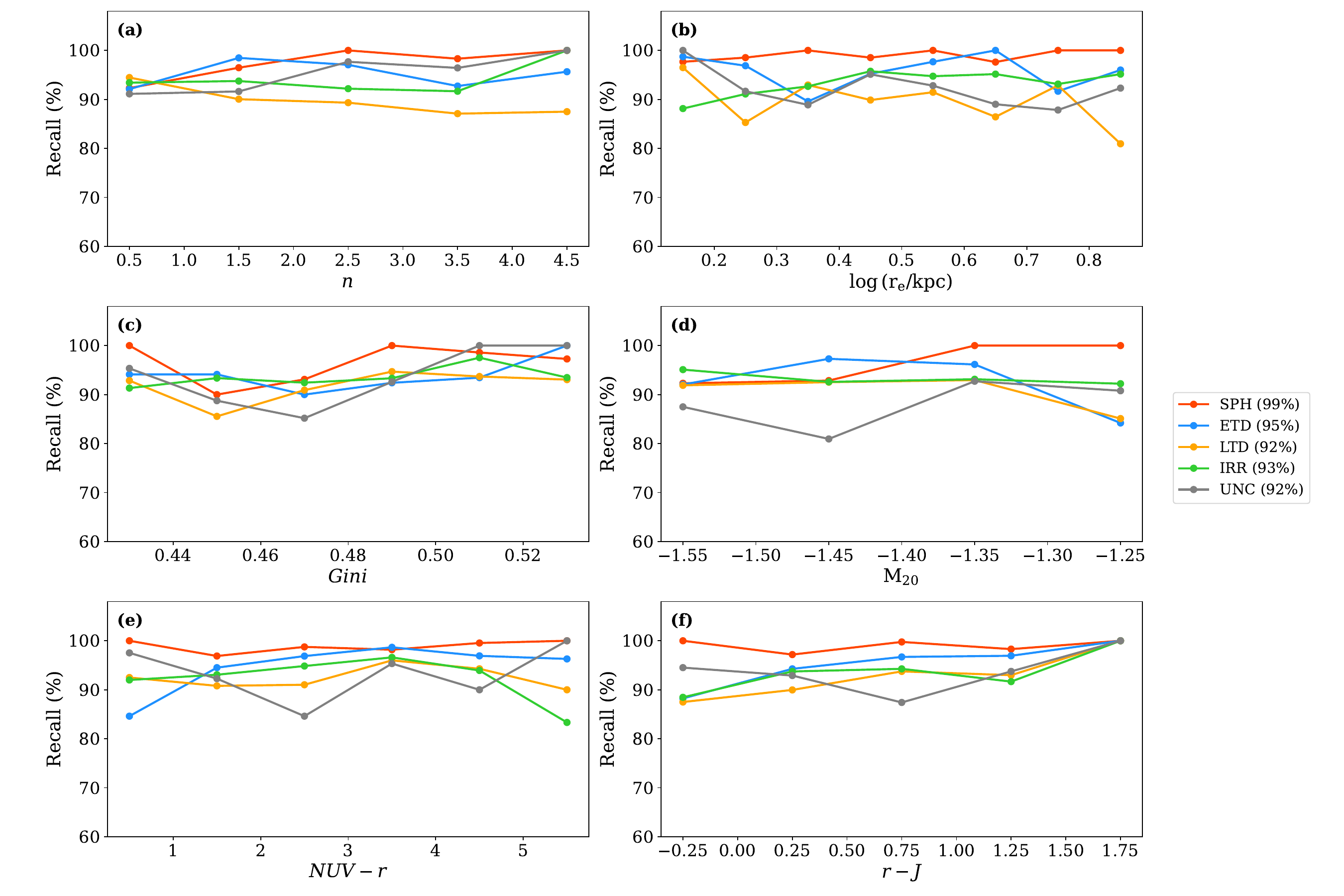}
    \caption{Same as Figure~\ref{fig:11} but for the recall metric. Similar to the results of precision, we find no significant difference or trend in the recall of each category in the parameter space, indicating the False Negatives do not rely on galaxy morphological and physical parameters. The recall of each category of galaxies in the parameter space of the validation set. The sequence and symbols are the same as Figure~\ref{fig:11}.  The model’s recall across different galaxy categories is independent of morphology and physical parameters, indicating that False Negatives show no significant dependence on these factors.
    }
    \label{fig:12}
\end{figure*}

\subsection{Overall morphological classification results and t-SNE Visualization}\label{sec:5.2}
By combining UML and SML, we successfully perform morphological classification of 99,806 galaxies in the COSMOS field, as shown in Table~\ref{tab5}. The final classification results comprise 15,326~SPH, 17,445~ETD, 20,931~LTD, 29,123~IRR, and 16,981~UNC. We then visualize the clustering results with the t-distributed stochastic neighbor embedding (t-SNE; \citealt{van+2008}), which projects high-dimensional data into a lower-dimensional space (e.g., 2D or 3D). As shown in Figure~\ref{fig:9}, t-SNE effectively maps the high-dimensional galaxy samples into a 2D plane while preserving both local similarities and global structural features.  t-SNE visualization allows us to qualitatively evaluate the classification performance of UML, SML, and the combined model. Figure~\ref{fig:9} demonstrates that all three methods produce well-separated class boundaries in the feature subspace: galaxies with similar features are accurately mapped to neighboring positions in the low-dimensional space, whereas galaxies with significant feature differences exhibit clear distributional separation. These results confirm the effectiveness of our classification approach.

\begin{deluxetable}{ccccccc}
\centerwidetable
\tablecaption{The number of galaxies classified into different types \label{tab5}}
\tablehead{\colhead{TYPE} & \colhead{SPH} & \colhead{ETD} & \colhead{LTD} & \colhead{IRR} & \colhead{UNC}
& \colhead{TOTAL}} 
\startdata  % 表格数据
UML &  8233 & 5322 & 6320 & 9468 & 7261 & 36,604\\
SML & 7093 & 12,123 & 14,611 & 19,655 & 9720 & 63,202\\
TOTAL & 15,326 & 17,445 & 20,931 & 29,123 & 16,981 & 99,806\\
\enddata
\end{deluxetable}

\subsection{Comparisons of Galaxy Properties}\label{sec:5.3}
To verify the consistency between the classification results and the known structural patterns of galaxies,
we analyze the classification results using the physical parameters of galaxies in Figure~\ref{fig:10}. To ensure the accuracy and reliability of the parameter measurements, the analysis is restricted to massive galaxies ($M_{*}>10^{9}~M_{\sun}$). Galaxies classified as UNC are excluded because their morphological parameters are difficult to measure due to the relatively low signal-to-noise ratios (S/N). In Figure~\ref{fig:10}, the SPH, ETD, LTD, and IRR categories are colored in red, blue, yellow, and green, respectively.

The panel (a) panel shows the distribution of galaxies in the rest-frame $NUV - r$ versus $r - J$ color space. The $NUV-J$ diagram is an effective diagnostic tool for distinguishing star-forming from quiescent galaxies; the wedge-shaped region is dominated by quiescent galaxies, while galaxies outside this region are considered star-forming (\citealt{Williams+2009}). Typically, ETD, LTD, and IRR galaxies primarily locate in the star-forming region, whereas most SPH galaxies are identified as quiescent.

The panel (b) panel presents the Gini coefficient ($G$) against the $M_{20}$ parameter (\citealt{Lotz+2004,Lotz+2008}). The Gini coefficient quantifies the non-uniformity of the surface brightness distribution, with higher values indicating stronger non-uniformity; $M_{20}$ is the second moment of the brightest 20\% of the galaxy’s light, with more negative values implying a higher degree of central concentration. 

The bottom panels show raincloud plots for the distributions of the Sérsic index $n$ (Panel \textbf{(c)}) and the effective radius $r_e$ (Panel \textbf{(d)}). Their definitions of these parameters are given in previous work (\citealt{Zhou+2022,Song+2024}).

Figure~\ref{fig:10} shows that our classification results are consistent with the expected evolutionary trend: along the sequence, from IRR to SPH types, (a) galaxies migrate from the star-forming region to the quiescent region, (b) their light distribution becomes more centrally concentrated, (c) galaxies gradually become more bulge-dominated, and (d) their morphology evolves from extended to more compact structures.

 We use the metrics of precision and recall to quantify the classification robustness of the GoogLeNet model across different parameter spaces and to investigate the relationship between FP and FN. Precision measures the accuracy of positive predictions, which increases as false positives decrease. Recall assesses the coverage of true positives prior to detection, which increases as false negatives decrease. Figures \ref{fig:11} and \ref{fig:12} show the classification precision and recall, respectively, for the five galaxy classes (SPH, ETD, LTD, IRR, UNC) on the validation set. The results show no significant correlation between these metrics and the morphological parameters of the galaxies. Both precision and recall remain above 90\% for all classes. Figures \ref{fig:11} and \ref{fig:12}  demonstrate that the distributions of true positives and false positives are independent of galactic characteristics. This confirms that the model delivers highly accurate and stable classification outputs across all categories, effectively avoiding biases introduced by specific sample properties.

\section{Summary and outlooks} \label{sec:6}
This paper presents a systematic analysis of the robustness of the hybrid machine learning framework \texttt{USmorph}, which integrates UML and SML for the automated classification of large-scale galaxy morphologies. The framework is validated with a sample of nearly 100,000 I-band galaxies ($0.2 < z < 1.2$, $I_{\mathrm{mag}} < 25$) drawn from the COSMOS field.

In the unsupervised learning phase, we verify that: (1) A CAE with two network layers, $5\times5$ convolutional kernels, and a 40-dimensional latent representation effectively suppresses background noise while maximally preserving key galactic structures. (2) The APCT is proven indispensable for achieving rotational invariance. APCT transforms rotational invariance into translational invariance, thereby conserving computational resources and enhancing the efficiency of downstream tasks.
(3) Bagging multi-clustering strategy yields high-purity consensus label classes. Setting the number of clusters to K=50 balances classification granularity with annotation efficiency and facilitates subsequent exploration of special objects (e.g., tidal tails, gravitational lenses, and little red dots) through subclass refinement.

In the supervised learning phase, the GoogLeNet model achieves a consistent overall classification accuracy exceeding 94\% across different dataset splits, demonstrating remarkable stability. Further validation involves t-SNE visualization and physical morphological parameters in the classification study. The results indicate that morphology classification based on \texttt{USmorph} exhibits distinct low-dimensional clustering boundaries in t-SNE visualization and maintains a high degree of consistency with the evolution of galaxy morphological parameter features. Additionally, both true positive and false positive samples displayed unbiased distributions within the parameter space, further substantiating the high reliability and stability of this classification method.

In future work, we will further enhance the automatic galaxy morphology classification framework by developing more effective large-scale model coding and dimensionality reduction techniques. We will also assess the stability and effectiveness of these algorithms (e.g., Ying et al. 2025, in preparation). We aim to improve the accuracy and generalization performance of deep learning models. This framework is anticipated to be applied to upcoming multi-band deep-field surveys, such as the imaging data expected from CSST, thereby advancing the study of galaxy morphology.

\section*{Data Availability}
Our code, dataset, and trained model weights can be found at \url{https://github.com/IAAA-246011/USmorph}.

\begin{acknowledgements}
This work is supported by the National Natural Science Foundation of China (NSFC) (grant Nos. 12233008 and 12573012, and Project No. 12503011), the National Key R\&D Program of China (grant No. 2023YFA1608100), the Strategic Priority Research Program of the Chinese Academy of Sciences (grant No. XDB0550200), the Cyrus Chun Ying Tang Foundations, and the 111 Project for “Observational and Theoretical Research on Dark Matter and Dark Energy” (B23042), and the China Manned Space Program with grant No. CMS-CSST-2025-A04. S.L. acknowledges the support from the Key Laboratory of Modern Astronomy and Astrophysics (Nanjing University) by the Ministry of Education. The numerical calculations in this paper have been done on the platform of the High Performance Computing of Anqing Normal University.

\end{acknowledgements}

\bibliography{ref}

@article{Prevot+1984,
  title = {The Typical Interstellar Extinction in the {{Small Magellanic Cloud}}.},
  author = {Prevot, M. L. and Lequeux, J. and Maurice, E. and Prevot, L. and {Rocca-Volmerange}, B.},
  year = {1984},
  journal = {Astronomy \& Astrophysics},
  volume = {132},
  pages = {389--392},
  adsurl = {https://ui.adsabs.harvard.edu/abs/1984A\&A...132..389P}
}

@ARTICLE{Lintott+2021,
    author = {{Lintott}, Chris and {Schawinski}, Kevin and {Bamford}, Steven and 
              {Slosar}, Anže and {Land}, Kate and {Thomas}, Daniel and 
              {Edmondson}, Edd and {Masters}, Karen and {Nichol}, Robert C. and 
              {Raddick}, M. Jordan and {Szalay}, Alex and {Andreescu}, Dan and 
              {Murray}, Phil and {Vandenberg}, Jan},
    title = "{Galaxy Zoo 1: Data Release of Morphological Classifications for Nearly 900,000 Galaxies}",
    journal = {\mnras},
    year = 2011,
    month = jan,
    volume = {410},
    number = {1},
    pages = {166-178},
    doi = {10.1111/j.1365-2966.2010.17432.x},
    archivePrefix = {arXiv},
    eprint = {1007.3265},
    primaryClass = {astro-ph.GA},
    adsurl = {https://ui.adsabs.harvard.edu/abs/2011MNRAS.410..166L}
}

@ARTICLE{Tohill+2024,
       author = {{Tohill}, C. and {Bamford}, S.~P. and {Conselice}, C.~J. and {Ferreira}, L. and {Harvey}, T. and {Adams}, N. and {Austin}, D.},
        title = "{A Robust Study of High-redshift Galaxies: Unsupervised Machine Learning for Characterizing Morphology with JWST up to z {\ensuremath{\sim}} 8}",
      journal = {\apj},
         year = 2024,
        month = feb,
       volume = {962},
       number = {2},
          eid = {164},
        pages = {164},
          doi = {10.3847/1538-4357/ad17b8},
archivePrefix = {arXiv},
       eprint = {2306.17225},
 primaryClass = {astro-ph.GA},
       adsurl = {https://ui.adsabs.harvard.edu/abs/2024ApJ...962..164T}
}

@article{Krizhevsky+2012,
  title={ImageNet classification with deep convolutional neural networks},
  author={Alex Krizhevsky and Ilya Sutskever and Geoffrey E. Hinton},
  journal={Communications of the ACM},
  year={2012},
  volume={60},
  pages={84 - 90},
  url={https://api.semanticscholar.org/CorpusID:195908774}
}

@article{Kingma+2014,
  title={Adam: A Method for Stochastic Optimization},
  author={Diederik P. Kingma and Jimmy Ba},
  journal={CoRR},
  year={2014},
  volume={abs/1412.6980},
  url={https://api.semanticscholar.org/CorpusID:6628106}
}

@INPROCEEDINGS{Deng+2020,
  author={Deng, Dingsheng},
  booktitle={2020 7th International Forum on Electrical Engineering and Automation (IFEEA)}, 
  title={DBSCAN Clustering Algorithm Based on Density}, 
  year={2020},
  volume={},
  number={},
  pages={949-953},
  doi={10.1109/IFEEA51475.2020.00199}}

@inproceedings{Sutskever+2013,
author = {Sutskever, Ilya and Martens, James and Dahl, George and Hinton, Geoffrey},
title = {On the importance of initialization and momentum in deep learning},
year = {2013},
publisher = {JMLR.org},
booktitle = {Proceedings of the 30th International Conference on International Conference on Machine Learning - Volume 28},
pages = {III–1139–III–1147},
location = {Atlanta, GA, USA},
series = {ICML'13}
}

@misc{Abell+2009,
      title={LSST Science Book, Version 2.0}, 
      author={LSST Science Collaboration and Paul A. Abell and Julius Allison and Scott F. Anderson and John R. Andrew and J. Roger P. Angel and Lee Armus and David Arnett and S. J. Asztalos and Tim S. Axelrod and Stephen Bailey and D. R. Ballantyne and Justin R. Bankert and Wayne A. Barkhouse and Jeffrey D. Barr and L. Felipe Barrientos and Aaron J. Barth and James G. Bartlett and Andrew C. Becker and Jacek Becla and Timothy C. Beers and Joseph P. Bernstein and Rahul Biswas and Michael R. Blanton and Joshua S. Bloom and John J. Bochanski and Pat Boeshaar and Kirk D. Borne and Marusa Bradac and W. N. Brandt and Carrie R. Bridge and Michael E. Brown and Robert J. Brunner and James S. Bullock and Adam J. Burgasser and James H. Burge and David L. Burke and Phillip A. Cargile and Srinivasan Chandrasekharan and George Chartas and Steven R. Chesley and You-Hua Chu and David Cinabro and Mark W. Claire and Charles F. Claver and Douglas Clowe and A. J. Connolly and Kem H. Cook and Jeff Cooke and Asantha Cooray and Kevin R. Covey and Christopher S. Culliton and Roelof de Jong and Willem H. de Vries and Victor P. Debattista and Francisco Delgado and Ian P. Dell'Antonio and Saurav Dhital and Rosanne Di Stefano and Mark Dickinson and Benjamin Dilday and S. G. Djorgovski and Gregory Dobler and Ciro Donalek and Gregory Dubois-Felsmann and Josef Durech and Ardis Eliasdottir and Michael Eracleous and Laurent Eyer and Emilio E. Falco and Xiaohui Fan and Christopher D. Fassnacht and Harry C. Ferguson and Yanga R. Fernandez and Brian D. Fields and Douglas Finkbeiner and Eduardo E. Figueroa and Derek B. Fox and Harold Francke and James S. Frank and Josh Frieman and Sebastien Fromenteau and Muhammad Furqan and Gaspar Galaz and A. Gal-Yam and Peter Garnavich and Eric Gawiser and John Geary and Perry Gee and Robert R. Gibson and Kirk Gilmore and Emily A. Grace and Richard F. Green and William J. Gressler and Carl J. Grillmair and Salman Habib and J. S. Haggerty and Mario Hamuy and Alan W. Harris and Suzanne L. Hawley and Alan F. Heavens and Leslie Hebb and Todd J. Henry and Edward Hileman and Eric J. Hilton and Keri Hoadley and J. B. Holberg and Matt J. Holman and Steve B. Howell and Leopoldo Infante and Zeljko Ivezic and Suzanne H. Jacoby and Bhuvnesh Jain and R and Jedicke and M. James Jee and J. Garrett Jernigan and Saurabh W. Jha and Kathryn V. Johnston and R. Lynne Jones and Mario Juric and Mikko Kaasalainen and Styliani and Kafka and Steven M. Kahn and Nathan A. Kaib and Jason Kalirai and Jeff Kantor and Mansi M. Kasliwal and Charles R. Keeton and Richard Kessler and Zoran Knezevic and Adam Kowalski and Victor L. Krabbendam and K. Simon Krughoff and Shrinivas Kulkarni and Stephen Kuhlman and Mark Lacy and Sebastien Lepine and Ming Liang and Amy Lien and Paulina Lira and Knox S. Long and Suzanne Lorenz and Jennifer M. Lotz and R. H. Lupton and Julie Lutz and Lucas M. Macri and Ashish A. Mahabal and Rachel Mandelbaum and Phil Marshall and Morgan May and Peregrine M. McGehee and Brian T. Meadows and Alan Meert and Andrea Milani and Christopher J. Miller and Michelle Miller and David Mills and Dante Minniti and David Monet and Anjum S. Mukadam and Ehud Nakar and Douglas R. Neill and Jeffrey A. Newman and Sergei Nikolaev and Martin Nordby and Paul O'Connor and Masamune Oguri and John Oliver and Scot S. Olivier and Julia K. Olsen and Knut Olsen and Edward W. Olszewski and Hakeem Oluseyi and Nelson D. Padilla and Alex Parker and Joshua Pepper and John R. Peterson and Catherine Petry and Philip A. Pinto and James L. Pizagno and Bogdan Popescu and Andrej Prsa and Veljko Radcka and M. Jordan Raddick and Andrew Rasmussen and Arne Rau and Jeonghee Rho and James E. Rhoads and Gordon T. Richards and Stephen T. Ridgway and Brant E. Robertson and Rok Roskar and Abhijit Saha and Ata Sarajedini and Evan Scannapieco and Terry Schalk and Rafe Schindler and Samuel Schmidt and Sarah Schmidt and Donald P. Schneider and German Schumacher and Ryan Scranton and Jacques Sebag and Lynn G. Seppala and Ohad Shemmer and Joshua D. Simon and M. Sivertz and Howard A. Smith and J. Allyn Smith and Nathan Smith and Anna H. Spitz and Adam Stanford and Keivan G. Stassun and Jay Strader and Michael A. Strauss and Christopher W. Stubbs and Donald W. Sweeney and Alex Szalay and Paula Szkody and Masahiro Takada and Paul Thorman and David E. Trilling and Virginia Trimble and Anthony Tyson and Richard Van Berg and Daniel Vanden Berk and Jake VanderPlas and Licia Verde and Bojan Vrsnak and Lucianne M. Walkowicz and Benjamin D. Wandelt and Sheng Wang and Yun Wang and Michael Warner and Risa H. Wechsler and Andrew A. West and Oliver Wiecha and Benjamin F. Williams and Beth Willman and David Wittman and Sidney C. Wolff and W. Michael Wood-Vasey and Przemek Wozniak and Patrick Young and Andrew Zentner and Hu Zhan},
      year={2009},
      eprint={0912.0201},
      archivePrefix={arXiv},
      primaryClass={astro-ph.IM},
      url={https://arxiv.org/abs/0912.0201}, 
}

@ARTICLE{Euclid+2025,
       author = {{Euclid Collaboration} and {Mellier}, Y. and {Abdurro'uf} and {Acevedo Barroso}, J.~A. and {Ach{\'u}carro}, A. and {Adamek}, J. and {Adam}, R. and {Addison}, G.~E. and {Aghanim}, N. and {Aguena}, M. and {Ajani}, V. and {Akrami}, Y. and {Al-Bahlawan}, A. and {Alavi}, A. and {Albuquerque}, I.~S. and {Alestas}, G. and {Alguero}, G. and {Allaoui}, A. and {Allen}, S.~W. and {Allevato}, V. and {Alonso-Tetilla}, A.~V. and {Altieri}, B. and {Alvarez-Candal}, A. and {Alvi}, S. and {Amara}, A. and {Amendola}, L. and {Amiaux}, J. and {Andika}, I.~T. and {Andreon}, S. and {Andrews}, A. and {Angora}, G. and {Angulo}, R.~E. and {Annibali}, F. and {Anselmi}, A. and {Anselmi}, S. and {Arcari}, S. and {Archidiacono}, M. and {Aric{\`o}}, G. and {Arnaud}, M. and {Arnouts}, S. and {Asgari}, M. and {Asorey}, J. and {Atayde}, L. and {Atek}, H. and {Atrio-Barandela}, F. and {Aubert}, M. and {Aubourg}, E. and {Auphan}, T. and {Auricchio}, N. and {Aussel}, B. and {Aussel}, H. and {Avelino}, P.~P. and {Avgoustidis}, A. and {Avila}, S. and {Awan}, S. and {Azzollini}, R. and {Baccigalupi}, C. and {Bachelet}, E. and {Bacon}, D. and {Baes}, M. and {Bagley}, M.~B. and {Bahr-Kalus}, B. and {Balaguera-Antolinez}, A. and {Balbinot}, E. and {Balcells}, M. and {Baldi}, M. and {Baldry}, I. and {Balestra}, A. and {Ballardini}, M. and {Ballester}, O. and {Balogh}, M. and {Ba{\~n}ados}, E. and {Barbier}, R. and {Bardelli}, S. and {Baron}, M. and {Barreiro}, T. and {Barrena}, R. and {Barriere}, J. -C. and {Barros}, B.~J. and {Barthelemy}, A. and {Bartolo}, N. and {Basset}, A. and {Battaglia}, P. and {Battisti}, A.~J. and {Baugh}, C.~M. and {Baumont}, L. and {Bazzanini}, L. and {Beaulieu}, J. -P. and {Beckmann}, V. and {Belikov}, A.~N. and {Bel}, J. and {Bellagamba}, F. and {Bella}, M. and {Bellini}, E. and {Benabed}, K. and {Bender}, R. and {Benevento}, G. and {Bennett}, C.~L. and {Benson}, K. and {Bergamini}, P. and {Bermejo-Climent}, J.~R. and {Bernardeau}, F. and {Bertacca}, D. and {Berthe}, M. and {Berthier}, J. and {Bethermin}, M. and {Beutler}, F. and {Bevillon}, C. and {Bhargava}, S. and {Bhatawdekar}, R. and {Bianchi}, D. and {Bisigello}, L. and {Biviano}, A. and {Blake}, R.~P. and {Blanchard}, A. and {Blazek}, J. and {Blot}, L. and {Bosco}, A. and {Bodendorf}, C. and {Boenke}, T. and {B{\"o}hringer}, H. and {Boldrini}, P. and {Bolzonella}, M. and {Bonchi}, A. and {Bonici}, M. and {Bonino}, D. and {Bonino}, L. and {Bonvin}, C. and {Bon}, W. and {Booth}, J.~T. and {Borgani}, S. and {Borlaff}, A.~S. and {Borsato}, E. and {Bose}, B. and {Botticella}, M.~T. and {Boucaud}, A. and {Bouche}, F. and {Boucher}, J.~S. and {Boutigny}, D. and {Bouvard}, T. and {Bouwens}, R. and {Bouy}, H. and {Bowler}, R.~A.~A. and {Bozza}, V. and {Bozzo}, E. and {Branchini}, E. and {Brando}, G. and {Brau-Nogue}, S. and {Brekke}, P. and {Bremer}, M.~N. and {Brescia}, M. and {Breton}, M. -A. and {Brinchmann}, J. and {Brinckmann}, T. and {Brockley-Blatt}, C. and {Brodwin}, M. and {Brouard}, L. and {Brown}, M.~L. and {Bruton}, S. and {Bucko}, J. and {Buddelmeijer}, H. and {Buenadicha}, G. and {Buitrago}, F. and {Burger}, P. and {Burigana}, C. and {Busillo}, V. and {Busonero}, D. and {Cabanac}, R. and {Cabayol-Garcia}, L. and {Cagliari}, M.~S. and {Caillat}, A. and {Caillat}, L. and {Calabrese}, M. and {Calabro}, A. and {Calderone}, G. and {Calura}, F. and {Camacho Quevedo}, B. and {Camera}, S. and {Campos}, L. and {Ca{\~n}as-Herrera}, G. and {Candini}, G.~P. and {Cantiello}, M. and {Capobianco}, V. and {Cappellaro}, E. and {Cappelluti}, N. and {Cappi}, A. and {Caputi}, K.~I. and {Cara}, C. and {Carbone}, C. and {Cardone}, V.~F. and {Carella}, E. and {Carlberg}, R.~G. and {Carle}, M. and {Carminati}, L. and {Caro}, F. and {Carrasco}, J.~M. and {Carretero}, J. and {Carrilho}, P. and {Carron Duque}, J. and {Carry}, B.},
        title = "{Euclid: I. Overview of the Euclid mission}",
      journal = {\aap},
         year = 2025,
        month = may,
       volume = {697},
          eid = {A1},
        pages = {A1},
          doi = {10.1051/0004-6361/202450810},
archivePrefix = {arXiv},
       eprint = {2405.13491},
 primaryClass = {astro-ph.CO},
       adsurl = {https://ui.adsabs.harvard.edu/abs/2025A&A...697A...1E}
}

@misc{CSST+2025,
      title={Introduction to the China Space Station Telescope (CSST)}, 
      author={CSST Collaboration and Yan Gong and Haitao Miao and Hu Zhan and Zhao-Yu Li and Jinyi Shangguan and Haining Li and Chao Liu and Xuefei Chen and Haibo Yuan and Jilin Zhou and Hui-Gen Liu and Cong Yu and Jianghui Ji and Zhaoxiang Qi and Jiacheng Liu and Zigao Dai and Xiaofeng Wang and Zhenya Zheng and Lei Hao and Jiangpei Dou and Yiping Ao and Zhenhui Lin and Kun Zhang and Wei Wang and Guotong Sun and Ran Li and Guoliang Li and Youhua Xu and Xinfeng Li and Shengyang Li and Peng Wu and Jiuxing Zhang and Bo Wang and Jinming Bai and Yi-Fu Cai and Zheng Cai and Kwan Chuen Chan and Jin Chang and Xiaodian Chen and Xuelei Chen and Yuqin Chen and Yun Chen and Wei Cui and Pu Du and Wenying Duan and Junhui Fan and LuLu Fan and Zhou Fan and Zuhui Fan and Taotao Fang and Jianning Fu and Liping Fu and Zhensen Fu and Jian Gao and Shenghong Gu and Yidong Gu and Qi Guo and Zhanwen Han and Zhiqi Huang and Luis C. Ho and Linhua Jiang and Yipeng Jing and Xi Kang and Xu Kong and Chengyuan Li and Di Li and Jing Li and Nan Li and Yang A. Li and Shilong Liao and Weipeng Lin and Fengshan Liu and Jifeng Liu and Xiangkun Liu and Ruiqing Mao and Shude Mao and Xianmin Meng and Xiaoying Pang and Xiyan Peng and Yingjie Peng and Huanyuan Shan and Juntai Shen and Shiyin Shen and Zhiqiang Shen and Sheng-Cai Shi and Yong Shi and Siyuan Tan and Hao Tian and Jianmin Wang and Jun-Xian Wang and Xin Wang and Yuting Wang and Hong Wu and Jingwen Wu and Xuebing Wu and Chun Xu and Xiang-Xiang Xue and Yongquan Xue and Ji Yang and Xiaohu Yang and Qijun Yao and Fangting Yuan and Zhen Yuan and Jun Zhang and Wei Zhang and Xin Zhang and Gang Zhao and Gongbo Zhao and Hongen Zhong and Jing Zhong and Liyong Zhou and Ying Zu},
      year={2025},
      eprint={2507.04618},
      archivePrefix={arXiv},
      primaryClass={astro-ph.IM},
      url={https://arxiv.org/abs/2507.04618}, 
}

@ARTICLE{S+1963,
       author = {{S{\'e}rsic}, J.~L.},
        title = "{Influence of the atmospheric and instrumental dispersion on the brightness distribution in a galaxy}",
      journal = {Boletin de la Asociacion Argentina de Astronomia La Plata Argentina},
         year = 1963,
        month = feb,
       volume = {6},
        pages = {41-43},
       adsurl = {https://ui.adsabs.harvard.edu/abs/1963BAAA....6...41S}
}

@INPROCEEDINGS{Deng+2009,
  author={Deng, Jia and Dong, Wei and Socher, Richard and Li, Li-Jia and Kai Li and Li Fei-Fei},
  booktitle={2009 IEEE Conference on Computer Vision and Pattern Recognition}, 
  title={ImageNet: A large-scale hierarchical image database}, 
  year={2009},
  volume={},
  number={},
  pages={248-255},
  doi={10.1109/CVPR.2009.5206848}}

@ARTICLE{sun+2018,
  author={Sun, Weiwei and Wang, Ruisheng},
  journal={IEEE Geoscience and Remote Sensing Letters}, 
  title={Fully Convolutional Networks for Semantic Segmentation of Very High Resolution Remotely Sensed Images Combined With DSM}, 
  year={2018},
  volume={15},
  number={3},
  pages={474-478},
  doi={10.1109/LGRS.2018.2795531}}

@ARTICLE{Williams+2009,
       author = {{Williams}, Rik J. and {Quadri}, Ryan F. and {Franx}, Marijn and {van Dokkum}, Pieter and {Labb{\'e}}, Ivo},
        title = "{Detection of Quiescent Galaxies in a Bicolor Sequence from Z = 0-2}",
      journal = {\apj},
         year = 2009,
        month = feb,
       volume = {691},
       number = {2},
        pages = {1879-1895},
          doi = {10.1088/0004-637X/691/2/1879},
archivePrefix = {arXiv},
       eprint = {0806.0625},
 primaryClass = {astro-ph},
       adsurl = {https://ui.adsabs.harvard.edu/abs/2009ApJ...691.1879W}
}

@article{Bourlard+1988,
author = {Bourlard, H. and Kamp, Y.},
title = {Auto-association by multilayer perceptrons and singular value decomposition},
year = {1988},
issue_date = {September 1988},
publisher = {Springer-Verlag},
address = {Berlin, Heidelberg},
volume = {59},
number = {4–5},
issn = {0340-1200},
url = {https://doi.org/10.1007/BF00332918},
doi = {10.1007/BF00332918},
journal = {Biol. Cybern.},
month = sep,
pages = {291–294},
numpages = {4}
}

@article{Kolmogorov+1933,
  title   = {Sulla determinazione empirica di una legge di distribuzione},
  author  = {Kolmogorov, Andrej Nikolajevi{\v{c}}},  
  journal = {Giornale dell'Istituto Italiano degli Attuari},
  year    = {1933},
  volume  = {4},
  pages   = {83--91},
  url     = {https://api.semanticscholar.org/CorpusID:222427298}
}

@ARTICLE{Comaniciu+2002,
  author={Comaniciu, D. and Meer, P.},
  journal={IEEE Transactions on Pattern Analysis and Machine Intelligence}, 
  title={Mean shift: a robust approach toward feature space analysis}, 
  year={2002},
  volume={24},
  number={5},
  pages={603-619},
  doi={10.1109/34.1000236}}

@article{Kolesnikov+2023,
    author = {Kolesnikov, I and Sampaio, V M and de Carvalho, R R and Conselice, C and Rembold, S B and Mendes, C L and Rosa, R R},
    title = {Unveiling galaxy morphology through an unsupervised-supervised hybrid approach},
    journal = {Monthly Notices of the Royal Astronomical Society},
    volume = {528},
    number = {1},
    pages = {82-107},
    year = {2023},
    month = {12},
    issn = {0035-8711},
    doi = {10.1093/mnras/stad3934},
    url = {https://doi.org/10.1093/mnras/stad3934},
    eprint = {https://academic.oup.com/mnras/article-pdf/528/1/82/55675427/stad3934.pdf},
}

@ARTICLE{su+2025,
       author = {{Su}, Yuguo and {Fang}, Guanwen and {Lu}, Shiying and {Lin}, Zesen},
        title = "{The impact of morphological quenching mechanisms on star formation activity at 0.2 < z < 1.2 in the COSMOS field}",
      journal = {\aap},
         year = 2025,
        month = jul,
       volume = {699},
          eid = {A184},
        pages = {A184},
          doi = {10.1051/0004-6361/202553693},
       adsurl = {https://ui.adsabs.harvard.edu/abs/2025A&A...699A.184S}
}

@article{Ilbert+2009,
doi = {10.1088/0004-637X/690/2/1236},
url = {https://dx.doi.org/10.1088/0004-637X/690/2/1236},
year = {2008},
month = {dec},
publisher = {The American Astronomical Society},
volume = {690},
number = {2},
pages = {1236},
author = {Ilbert, O. and Capak, P. and Salvato, M. and Aussel, H. and McCracken, H. J. and Sanders, D. B. and Scoville, N. and Kartaltepe, J. and Arnouts, S. and Floc'h, E. Le and Mobasher, B. and Taniguchi, Y. and Lamareille, F. and Leauthaud, A. and Sasaki, S. and Thompson, D. and Zamojski, M. and Zamorani, G. and Bardelli, S. and Bolzonella, M. and Bongiorno, A. and Brusa, M. and Caputi, K. I. and Carollo, C. M. and Contini, T. and Cook, R. and Coppa, G. and Cucciati, O. and de la Torre, S. and de Ravel, L. and Franzetti, P. and Garilli, B. and Hasinger, G. and Iovino, A. and Kampczyk, P. and Kneib, J.-P. and Knobel, C. and Kovac, K. and Le Borgne, J. F. and Le Brun, V. and Fèvre, O. Le and Lilly, S. and Looper, D. and Maier, C. and Mainieri, V. and Mellier, Y. and Mignoli, M. and Murayama, T. and Pellò, R. and Peng, Y. and Pérez-Montero, E. and Renzini, A. and Ricciardelli, E. and Schiminovich, D. and Scodeggio, M. and Shioya, Y. and Silverman, J. and Surace, J. and Tanaka, M. and Tasca, L. and Tresse, L. and Vergani, D. and Zucca, E.},
title = {COSMOS PHOTOMETRIC REDSHIFTS WITH 30-BANDS FOR 2-deg2},
journal = {The Astrophysical Journal}
}

@article{Bruzual+2003,
    author = {Bruzual, G. and Charlot, S.},
    title = "{Stellar population synthesis at the resolution of 2003}",
    journal = {Monthly Notices of the Royal Astronomical Society},
    volume = {344},
    number = {4},
    pages = {1000-1028},
    year = {2003},
    month = {10},
    issn = {0035-8711},
    doi = {10.1046/j.1365-8711.2003.06897.x},
    url = {https://doi.org/10.1046/j.1365-8711.2003.06897.x},
    eprint = {https://academic.oup.com/mnras/article-pdf/344/4/1000/2908334/344-4-1000.pdf},
}

@ARTICLE{Hocking+2018,
       author = {{Hocking}, Alex and {Geach}, James E. and {Sun}, Yi and {Davey}, Neil},
        title = "{An automatic taxonomy of galaxy morphology using unsupervised machine learning}",
      journal = {\mnras},
         year = 2018,
        month = jan,
       volume = {473},
       number = {1},
        pages = {1108-1129},
          doi = {10.1093/mnras/stx2351},
archivePrefix = {arXiv},
       eprint = {1709.05834},
 primaryClass = {astro-ph.IM},
       adsurl = {https://ui.adsabs.harvard.edu/abs/2018MNRAS.473.1108H}
}

@ARTICLE{Ilbert+2006,
       author = {{Ilbert}, O. and {Arnouts}, S. and {McCracken}, H.~J. and {Bolzonella}, M. and {Bertin}, E. and {Le F{\`e}vre}, O. and {Mellier}, Y. and {Zamorani}, G. and {Pell{\`o}}, R. and {Iovino}, A. and {Tresse}, L. and {Le Brun}, V. and {Bottini}, D. and {Garilli}, B. and {Maccagni}, D. and {Picat}, J.~P. and {Scaramella}, R. and {Scodeggio}, M. and {Vettolani}, G. and {Zanichelli}, A. and {Adami}, C. and {Bardelli}, S. and {Cappi}, A. and {Charlot}, S. and {Ciliegi}, P. and {Contini}, T. and {Cucciati}, O. and {Foucaud}, S. and {Franzetti}, P. and {Gavignaud}, I. and {Guzzo}, L. and {Marano}, B. and {Marinoni}, C. and {Mazure}, A. and {Meneux}, B. and {Merighi}, R. and {Paltani}, S. and {Pollo}, A. and {Pozzetti}, L. and {Radovich}, M. and {Zucca}, E. and {Bondi}, M. and {Bongiorno}, A. and {Busarello}, G. and {de La Torre}, S. and {Gregorini}, L. and {Lamareille}, F. and {Mathez}, G. and {Merluzzi}, P. and {Ripepi}, V. and {Rizzo}, D. and {Vergani}, D.},
        title = "{Accurate photometric redshifts for the CFHT legacy survey calibrated using the VIMOS VLT deep survey}",
      journal = {\aap},
         year = 2006,
        month = oct,
       volume = {457},
       number = {3},
        pages = {841-856},
          doi = {10.1051/0004-6361:20065138},
archivePrefix = {arXiv},
       eprint = {astro-ph/0603217},
 primaryClass = {astro-ph},
       adsurl = {https://ui.adsabs.harvard.edu/abs/2006A&A...457..841I}
}

@ARTICLE{Conselice+2003,
       author = {{Conselice}, Christopher J.},
        title = "{The Relationship between Stellar Light Distributions of Galaxies and Their Formation Histories}",
      journal = {\apjs},
         year = 2003,
        month = jul,
       volume = {147},
       number = {1},
        pages = {1-28},
          doi = {10.1086/375001},
archivePrefix = {arXiv},
       eprint = {astro-ph/0303065},
 primaryClass = {astro-ph},
       adsurl = {https://ui.adsabs.harvard.edu/abs/2003ApJS..147....1C}
}

@article{Massey+2009,
    author = {Massey, Richard and Stoughton, Chris and Leauthaud, Alexie and Rhodes, Jason and Koekemoer, Anton and Ellis, Richard and Shaghoulian, Edgar},
    title = "{Pixel-based correction for Charge Transfer Inefficiency in the Hubble Space Telescope Advanced Camera for Surveys}",
    journal = {Monthly Notices of the Royal Astronomical Society},
    volume = {401},
    number = {1},
    pages = {371-384},
    year = {2009},
    month = {12},
    issn = {0035-8711},
    doi = {10.1111/j.1365-2966.2009.15638.x},
    url = {https://doi.org/10.1111/j.1365-2966.2009.15638.x},
    eprint = {https://academic.oup.com/mnras/article-pdf/401/1/371/18581537/mnras0401-0371.pdf}
}

@ARTICLE{Stoughton+2002,
       author = {{Stoughton}, Chris and {Lupton}, Robert H. and {Bernardi}, Mariangela and {Blanton}, Michael R. and {Burles}, Scott and {Castander}, Francisco J. and {Connolly}, A.~J. and {Eisenstein}, Daniel J. and {Frieman}, Joshua A. and {Hennessy}, G.~S. and {Hindsley}, Robert B. and {Ivezi{\'c}}, {\v{Z}}eljko and {Kent}, Stephen and {Kunszt}, Peter Z. and {Lee}, Brian C. and {Meiksin}, Avery and {Munn}, Jeffrey A. and {Newberg}, Heidi Jo and {Nichol}, R.~C. and {Nicinski}, Tom and {Pier}, Jeffrey R. and {Richards}, Gordon T. and {Richmond}, Michael W. and {Schlegel}, David J. and {Smith}, J. Allyn and {Strauss}, Michael A. and {SubbaRao}, Mark and {Szalay}, Alexander S. and {Thakar}, Aniruddha R. and {Tucker}, Douglas L. and {Vanden Berk}, Daniel E. and {Yanny}, Brian and {Adelman}, Jennifer K. and {Anderson}, Jr., John E. and {Anderson}, Scott F. and {Annis}, James and {Bahcall}, Neta A. and {Bakken}, J.~A. and {Bartelmann}, Matthias and {Bastian}, Steven and {Bauer}, Amanda and {Berman}, Eileen and {B{\"o}hringer}, Hans and {Boroski}, William N. and {Bracker}, Steve and {Briegel}, Charlie and {Briggs}, John W. and {Brinkmann}, J. and {Brunner}, Robert and {Carey}, Larry and {Carr}, Michael A. and {Chen}, Bing and {Christian}, Damian and {Colestock}, Patrick L. and {Crocker}, J.~H. and {Csabai}, Istv{\'a}n and {Czarapata}, Paul C. and {Dalcanton}, Julianne and {Davidsen}, Arthur F. and {Davis}, John Eric and {Dehnen}, Walter and {Dodelson}, Scott and {Doi}, Mamoru and {Dombeck}, Tom and {Donahue}, Megan and {Ellman}, Nancy and {Elms}, Brian R. and {Evans}, Michael L. and {Eyer}, Laurent and {Fan}, Xiaohui and {Federwitz}, Glenn R. and {Friedman}, Scott and {Fukugita}, Masataka and {Gal}, Roy and {Gillespie}, Bruce and {Glazebrook}, Karl and {Gray}, Jim and {Grebel}, Eva K. and {Greenawalt}, Bruce and {Greene}, Gretchen and {Gunn}, James E. and {de Haas}, Ernst and {Haiman}, Zolt{\'a}n and {Haldeman}, Merle and {Hall}, Patrick B. and {Hamabe}, Masaru and {Hansen}, Brad and {Harris}, Frederick H. and {Harris}, Hugh and {Harvanek}, Michael and {Hawley}, Suzanne L. and {Hayes}, J.~J.~E. and {Heckman}, Timothy M. and {Helmi}, Amina and {Henden}, Arne and {Hogan}, Craig J. and {Hogg}, David W. and {Holmgren}, Donald J. and {Holtzman}, Jon and {Huang}, Chih-Hao and {Hull}, Charles and {Ichikawa}, Shin-Ichi and {Ichikawa}, Takashi and {Johnston}, David E. and {Kauffmann}, Guinevere and {Kim}, Rita S.~J. and {Kimball}, Tim and {Kinney}, E. and {Klaene}, Mark and {Kleinman}, S.~J. and {Klypin}, Anatoly and {Knapp}, G.~R. and {Korienek}, John and {Krolik}, Julian and {Kron}, Richard G. and {Krzesi{\'n}ski}, Jurek and {Lamb}, D.~Q. and {Leger}, R. French and {Limmongkol}, Siriluk and {Lindenmeyer}, Carl and {Long}, Daniel C. and {Loomis}, Craig and {Loveday}, Jon and {MacKinnon}, Bryan and {Mannery}, Edward J. and {Mantsch}, P.~M. and {Margon}, Bruce and {McGehee}, Peregrine and {McKay}, Timothy A. and {McLean}, Brian and {Menou}, Kristen and {Merelli}, Aronne and {Mo}, H.~J. and {Monet}, David G. and {Nakamura}, Osamu and {Narayanan}, Vijay K. and {Nash}, Thomas and {Neilsen}, Jr., Eric H. and {Newman}, Peter R. and {Nitta}, Atsuko and {Odenkirchen}, Michael and {Okada}, Norio and {Okamura}, Sadanori and {Ostriker}, Jeremiah P. and {Owen}, Russell and {Pauls}, A. George and {Peoples}, John and {Peterson}, R.~S. and {Petravick}, Donald and {Pope}, Adrian and {Pordes}, Ruth and {Postman}, Marc and {Prosapio}, Angela and {Quinn}, Thomas R. and {Rechenmacher}, Ron and {Rivetta}, Claudio H. and {Rix}, Hans-Walter and {Rockosi}, Constance M. and {Rosner}, Robert and {Ruthmansdorfer}, Kurt and {Sandford}, Dale and {Schneider}, Donald P. and {Scranton}, Ryan and {Sekiguchi}, Maki and {Sergey}, Gary and {Sheth}, Ravi and {Shimasaku}, Kazuhiro and {Smee}, Stephen and {Snedden}, Stephanie A. and {Stebbins}, Albert and {Stubbs}, Christopher and {Szapudi}, Istv{\'a}n and {Szkody}, Paula and {Szokoly}, Gyula P. and {Tabachnik}, Serge and {Tsvetanov}, Zlatan and {Uomoto}, Alan and {Vogeley}, Michael S. and {Voges}, Wolfgang and {Waddell}, Patrick and {Walterbos}, Ren{\'e} and {Wang}, Shu-i. and {Watanabe}, Masaru and {Weinberg}, David H. and {White}, Richard L. and {White}, Simon D.~M. and {Wilhite}, Brian and {Wolfe}, David and {Yasuda}, Naoki and {York}, Donald G. and {Zehavi}, Idit and {Zheng}, Wei},
        title = "{Sloan Digital Sky Survey: Early Data Release}",
      journal = {\aj},
         year = 2002,
        month = jan,
       volume = {123},
       number = {1},
        pages = {485-548},
          doi = {10.1086/324741},
       adsurl = {https://ui.adsabs.harvard.edu/abs/2002AJ....123..485S}
}

@article{Weaver+2022,
  author = {{Weaver}, J.~R. and {Kauffmann}, O.~B. and {Ilbert}, O. and {McCracken}, H.~J. and {Moneti}, A. and {Toft}, S. and {Brammer}, G. and {Shuntov}, M. and {Davidzon}, I. and {Hsieh}, B.~C. and {Laigle}, C. and {Anastasiou}, A. and {Jespersen}, C.~K. and {Vinther}, J. and {Capak}, P. and {Casey}, C.~M. and {McPartland}, C.~J.~R. and {Milvang-Jensen}, B. and {Mobasher}, B. and {Sanders}, D.~B. and {Zalesky}, L. and {Arnouts}, S. and {Aussel}, H. and {Dunlop}, J.~S. and {Faisst}, A. and {Franx}, M. and {Furtak}, L.~J. and {Fynbo}, J.~P.~U. and {Gould}, K.~M.~L. and {Greve}, T.~R. and {Gwyn}, S. and {Kartaltepe}, J.~S. and {Kashino}, D. and {Koekemoer}, A.~M. and {Kokorev}, V. and {Le Fèvre}, O. and {Lilly}, S. and {Masters}, D. and {Magdis}, G. and {Mehta}, V. and {Peng}, Y. and {Riechers}, D.~A. and {Salvato}, M. and {Sawicki}, M. and {Scarlata}, C. and {Scoville}, N. and {Shirley}, R. and {Silverman}, J.~D. and {Sneppen}, A. and {Smolčić}, V. and {Steinhardt}, C. and {Stern}, D. and {Tanaka}, M. and {Taniguchi}, Y. and {Teplitz}, H.~I. and {Vaccari}, M. and {Wang}, W.-H. and {Zamorani}, G.},
  title = {COSMOS2020: A Panchromatic View of the Universe to z ∼ 10 from Two Complementary Catalogs},
  journal = {The Astrophysical Journal Supplement Series},
  year = {2022},
  month = {jan},
  volume = {258},
  number = {1},
  pages = {11},
  doi = {10.3847/1538-4365/ac3078},
  url = {https://dx.doi.org/10.3847/1538-4365/ac3078}
}

@ARTICLE{Oke+1983,
       author = {{Oke}, J.~B. and {Gunn}, J.~E.},
        title = "{Secondary standard stars for absolute spectrophotometry.}",
      journal = {\apj},
         year = 1983,
        month = mar,
       volume = {266},
        pages = {713-717},
          doi = {10.1086/160817},
       adsurl = {https://ui.adsabs.harvard.edu/abs/1983ApJ...266..713O}
}

@article{Chabrier+2003,
doi = {10.1086/376392},
url = {https://dx.doi.org/10.1086/376392},
year = {2003},
month = {jul},
publisher = {The University of Chicago Press},
volume = {115},
number = {809},
pages = {763},
author = {Gilles Chabrier},
title = {Galactic Stellar and Substellar Initial Mass Function1},
journal = {Publications of the Astronomical Society of the Pacific},
}

@article{Zhang+2024,
doi = {10.1088/1674-4527/ad6fe6},
url = {https://dx.doi.org/10.1088/1674-4527/ad6fe6},
year = {2024},
month = {sep},
publisher = {National Astromonical Observatories, CAS and IOP Publishing},
volume = {24},
number = {9},
pages = {095012},
author = {Shiliang Zhang and Guanwen Fang and Jie Song and Ran Li and Yizhou Gu and Zesen Lin and Chichun Zhou and Yao Dai and Xu Kong},
title = {Preparation for CSST: Star-galaxy Classification using a Rotationally Invariant Supervised Machine Learning Method},
journal = {Research in Astronomy and Astrophysics}
}

@article{Kauffmann+2004,
  title = {The Environmental Dependence of the Relations between Stellar Mass, Structure, Star Formation and Nuclear Activity in Galaxies: {{Galaxy}} Structure, Star Formation and Nuclear Activity},
  shorttitle = {The Environmental Dependence of the Relations between Stellar Mass, Structure, Star Formation and Nuclear Activity in Galaxies},
  author = {Kauffmann, Guinevere and White, Simon D. M. and Heckman, Timothy M. and M{\'e}nard, Brice and Brinchmann, Jarle and Charlot, St{\'e}phane and Tremonti, Christy and Brinkmann, Jon},
  year = {2004},
  journal = {Monthly Notices of the Royal Astronomical Society},
  volume = {353},
  number = {3},
  pages = {713--731},
  doi = {10.1111/j.1365-2966.2004.08117.x},
  urldate = {2022-08-26},
  langid = {english}
}

@article{omand+2014,
  title = {The Connection between Galaxy Structure and Quenching Efficiency},
  author = {Omand, Conor M. B. and Balogh, Michael L. and Poggianti, Bianca M.},
  year = {2014},
  journal = {Monthly Notices of the Royal Astronomical Society},
  volume = {440},
  number = {1},
  pages = {843--858},
  doi = {10.1093/mnras/stu331},
  urldate = {2023-07-11},
  langid = {english}
}

@article{schawinski+2014,
  title = {The Green Valley Is a Red Herring: {{Galaxy Zoo}} Reveals Two Evolutionary Pathways towards Quenching of Star Formation in Early- and Late-Type Galaxies\ding{72}},
  shorttitle = {The Green Valley Is a Red Herring},
  author = {Schawinski, Kevin and Urry, C. Megan and Simmons, Brooke D. and Fortson, Lucy and Kaviraj, Sugata and Keel, William C. and Lintott, Chris J. and Masters, Karen L. and Nichol, Robert C. and Sarzi, Marc and Skibba, Ramin and Treister, Ezequiel and Willett, Kyle W. and Wong, O. Ivy and Yi, Sukyoung K.},
  year = {2014},
  journal = {Monthly Notices of the Royal Astronomical Society},
  volume = {440},
  number = {1},
  pages = {889--907},
  doi = {10.1093/mnras/stu327},
  urldate = {2022-08-26},
  langid = {english}
}

@article{kawinwanichakij+2017,
  title = {Effect of {{Local Environment}} and {{Stellar Mass}} on {{Galaxy Quenching}} and {{Morphology}} at 0.5 {$<$} z {$<$} 2.0},
  author = {Kawinwanichakij, Lalitwadee and Papovich, Casey and Quadri, Ryan F. and Glazebrook, Karl and Kacprzak, Glenn G. and Allen, Rebecca J. and Bell, Eric F. and Croton, Darren J. and Dekel, Avishai and Ferguson, Henry C. and Forrest, Ben and Grogin, Norman A. and Guo, Yicheng and Kocevski, Dale D. and Koekemoer, Anton M. and Labb{\'e}, Ivo and Lucas, Ray A. and Nanayakkara, Themiya and Spitler, Lee R. and Straatman, Caroline M. S. and Tran, Kim-Vy H. and Tomczak, Adam and van Dokkum, Pieter},
  year = {2017},
  journal = {The Astrophysical Journal},
  volume = {847},
  number = {2},
  pages = {134},
  doi = {10.3847/1538-4357/aa8b75},
  urldate = {2021-12-06},
  langid = {english}
}

@INPROCEEDINGS{yao+2019,
  author={Yao, Xiwen and Feng, Xiaoxu and Cheng, Gong and Han, Junwei and Guo, Lei},
  booktitle={IGARSS 2019 - 2019 IEEE International Geoscience and Remote Sensing Symposium}, 
  title={Rotation-Invariant Latent Semantic Representation Learning for Object Detection in VHR Optical Remote Sensing Images}, 
  year={2019},
  volume={},
  number={},
  pages={1382-1385},
  doi={10.1109/IGARSS.2019.8899285}}

@article{Scoville+2007,
   title={The Cosmic Evolution Survey (COSMOS): Overview},
   volume={172},
   ISSN={1538-4365},
   url={http://dx.doi.org/10.1086/516585},
   DOI={10.1086/516585},
   number={1},
   journal={The Astrophysical Journal Supplement Series},
   publisher={American Astronomical Society},
   author={Scoville, N. and Aussel, H. and Brusa, M. and Capak, P. and Carollo, C. M. and Elvis, M. and Giavalisco, M. and Guzzo, L. and Hasinger, G. and Impey, C. and Kneib, J.‐P. and LeFevre, O. and Lilly, S. J. and Mobasher, B. and Renzini, A. and Rich, R. M. and Sanders, D. B. and Schinnerer, E. and Schminovich, D. and Shopbell, P. and Taniguchi, Y. and Tyson, N. D.},
   year={2007},
   month=sep, pages={1–8} }

@article{Dieleman+2015,
   title={Rotation-invariant convolutional neural networks for galaxy morphology prediction},
   volume={450},
   ISSN={0035-8711},
   url={http://dx.doi.org/10.1093/mnras/stv632},
   DOI={10.1093/mnras/stv632},
   number={2},
   journal={Monthly Notices of the Royal Astronomical Society},
   publisher={Oxford University Press (OUP)},
   author={Dieleman, Sander and Willett, Kyle W. and Dambre, Joni},
   year={2015},
   month=apr, pages={1441–1459} }

@INPROCEEDINGS{Fernando+2024,
  author={Fernando, Thrinith and Rathnayake, Samadhi and Dissanayaka, Kapila},
  booktitle={2024 6th International Conference on Advancements in Computing (ICAC)}, 
  title={Galaxy Morphology Classification Based on VGG19 Deep Convolutional Neural Network}, 
  year={2024},
  volume={},
  number={},
  pages={229-234},
  doi={10.1109/ICAC64487.2024.10851157}}

@article{Zhou+2022,
   title={Automatic Morphological Classification of Galaxies: Convolutional Autoencoder and Bagging-based Multiclustering Model},
   volume={163},
   ISSN={1538-3881},
   url={http://dx.doi.org/10.3847/1538-3881/ac4245},
   DOI={10.3847/1538-3881/ac4245},
   number={2},
   journal={The Astronomical Journal},
   publisher={American Astronomical Society},
   author={Zhou, ChiChun and Gu, Yizhou and Fang, Guanwen and Lin, Zesen},
   year={2022},
   month=jan, pages={86} }

@article{Fang+2023,
doi = {10.3847/1538-3881/aca1a6},
url = {https://dx.doi.org/10.3847/1538-3881/aca1a6},
year = {2023},
month = {jan},
publisher = {The American Astronomical Society},
volume = {165},
number = {2},
pages = {35},
author = {GuanWen Fang and Shuo Ba and Yizhou Gu and Zesen Lin and Yuejie Hou and Chenxin Qin and Chichun Zhou and Jun Xu and Yao Dai and Jie Song and Xu Kong},
title = {Automatic Classification of Galaxy Morphology: A Rotationally-invariant Supervised Machine-learning Method Based on the Unsupervised Machine-learning Data Set},
journal = {The Astronomical Journal}
}

@article{van+2008,
title = "Visualizing High-Dimensional Data Using t-SNE",
author = "{van der Maaten}, L.J.P. and G.E. Hinton",
note = "Pagination: 27",
year = "2008",
language = "English",
volume = "9",
pages = "2579--2605",
journal = "Journal of Machine Learning Research",
issn = "1532-4435",
publisher = "Microtome Publishing",
number = "nov",
}

@article{Bramme+2008,
doi = {10.1086/591786},
url = {https://dx.doi.org/10.1086/591786},
year = {2008},
month = {oct},
publisher = {},
volume = {686},
number = {2},
pages = {1503},
author = {Gabriel B. Brammer and Pieter G. van Dokkum and Paolo Coppi},
title = {EAZY: A Fast, Public Photometric Redshift Code},
journal = {The Astrophysical Journal}
}

@article{Calzetti+2000,
doi = {10.1086/308692},
url = {https://dx.doi.org/10.1086/308692},
year = {2000},
month = {apr},
publisher = {},
volume = {533},
number = {2},
pages = {682},
author = {Daniela Calzetti and Lee Armus and Ralph C. Bohlin and Anne L. Kinney and Jan Koornneef and Thaisa Storchi-Bergmann},
title = {The Dust Content and Opacity of Actively
Star-forming Galaxies*},
journal = {The Astrophysical Journal}
}

@article{Lotz+2004,
doi = {10.1086/421849},
url = {https://dx.doi.org/10.1086/421849},
year = {2004},
month = {jul},
publisher = {},
volume = {128},
number = {1},
pages = {163},
author = {Jennifer M. Lotz and Joel Primack and Piero Madau},
title = {A New Nonparametric Approach to Galaxy Morphological Classification},
journal = {The Astronomical Journal}
}

@article{Lotz+2008,
doi = {10.1086/523659},
url = {https://dx.doi.org/10.1086/523659},
year = {2008},
month = {jan},
publisher = {},
volume = {672},
number = {1},
pages = {177},
author = {Jennifer M. Lotz and M. Davis and S. M. Faber and P. Guhathakurta and S. Gwyn and J. Huang and D. C. Koo and E. Le Floc’h and Lihwai Lin and J. Newman and K. Noeske and C. Papovich and C. N. A. Willmer and A. Coil and C. J. Conselice and M. Cooper and A. M. Hopkins and A. Metevier and J. Primack and G. Rieke and B. J. Weiner},
title = {The Evolution of Galaxy Mergers and Morphology at z &lt; 1.2 in the Extended Groth Strip},
journal = {The Astrophysical Journal}
}

@article{Davis+2014,
doi = {10.1088/0004-637X/790/2/87},
url = {https://dx.doi.org/10.1088/0004-637X/790/2/87},
year = {2014},
month = {jul},
publisher = {The American Astronomical Society},
volume = {790},
number = {2},
pages = {87},
author = {Davis, Darren R. and Hayes, Wayne B.},
title = {SpArcFiRe: SCALABLE AUTOMATED DETECTION OF SPIRAL GALAXY ARM SEGMENTS},
journal = {The Astrophysical Journal}
}

@article{Conselice+2000,
doi = {10.1086/308300},
url = {https://dx.doi.org/10.1086/308300},
year = {2000},
month = {feb},
publisher = {},
volume = {529},
number = {2},
pages = {886},
author = {Christopher J. Conselice and Matthew A. Bershady and Anna Jangren},
title = {The Asymmetry of Galaxies: Physical Morphology for Nearby and High-Redshift Galaxies},
journal = {The Astrophysical Journal}
}

@article{Wei+2024,
doi = {10.3847/1538-3881/ad8632},
url = {https://dx.doi.org/10.3847/1538-3881/ad8632},
year = {2024},
month = {nov},
publisher = {The American Astronomical Society},
volume = {168},
number = {6},
pages = {264},
author = {Wei, Junye and Xu, Ye and Lin, Zehao and Hao, Chaojie and Li, Yingjie and Liu, Dejian and Bian, Shuaibo},
title = {A New Statistical Analysis of the Morphology of Spiral Galaxies},
journal = {The Astronomical Journal}
}

@ARTICLE{Ho+2019,
       author = {{Ho}, Stephanie H. and {Martin}, Crystal L. and {Turner}, Monica L.},
        title = "{How Gas Accretion Feeds Galactic Disks}",
      journal = {\apj},
         year = 2019,
        month = apr,
       volume = {875},
       number = {1},
          eid = {54},
        pages = {54},
          doi = {10.3847/1538-4357/ab0ec2},
archivePrefix = {arXiv},
       eprint = {1903.06840},
 primaryClass = {astro-ph.GA},
       adsurl = {https://ui.adsabs.harvard.edu/abs/2019ApJ...875...54H}
}

@ARTICLE{ren+2023,
       author = {{Ren}, Jian and {Li}, Nan and {Liu}, F.~S. and {Cui}, Qifan and {Fu}, Mingxiang and {Zheng}, Xian Zhong},
        title = "{Revisiting Galaxy Evolution in Morphology in the Cosmic Evolution Survey Field (COSMOS-ReGEM). I. Merging Galaxies}",
      journal = {\apj},
         year = 2023,
        month = nov,
       volume = {958},
       number = {1},
          eid = {96},
        pages = {96},
          doi = {10.3847/1538-4357/acfeee},
archivePrefix = {arXiv},
       eprint = {2309.16531},
 primaryClass = {astro-ph.GA},
       adsurl = {https://ui.adsabs.harvard.edu/abs/2023ApJ...958...96R}
}

@ARTICLE{Sofue+2021,
  author={Sofue, Yoshiaki and Kataoka, Jun},
  journal={Monthly Notices of the Royal Astronomical Society}, 
  title={Interaction of the galactic-centre super bubbles with the gaseous disc}, 
  year={2021},
  volume={506},
  number={2},
  pages={2170-2180},
  doi={10.1093/mnras/stab1857}}

@article{Lintott+2008,
    author = {Lintott, Chris J. and Schawinski, Kevin and Slosar, Anže and Land, Kate and Bamford, Steven and Thomas, Daniel and Raddick, M. Jordan and Nichol, Robert C. and Szalay, Alex and Andreescu, Dan and Murray, Phil and Vandenberg, Jan},
    title = {Galaxy Zoo: morphologies derived from visual inspection of galaxies from the Sloan Digital Sky Survey*},
    journal = {Monthly Notices of the Royal Astronomical Society},
    volume = {389},
    number = {3},
    pages = {1179-1189},
    year = {2008},
    month = {09},
    issn = {0035-8711},
    doi = {10.1111/j.1365-2966.2008.13689.x},
    url = {https://doi.org/10.1111/j.1365-2966.2008.13689.x},
    eprint = {https://academic.oup.com/mnras/article-pdf/389/3/1179/3325962/mnras0389-1179.pdf},
}

@article{Balcells+2003,
doi = {10.1086/367783},
url = {https://dx.doi.org/10.1086/367783},
year = {2002},
month = {dec},
publisher = {},
volume = {582},
number = {2},
pages = {L79},
author = {Balcells, Marc and Graham, Alister W. and Domínguez-Palmero, Lilian and Peletier, Reynier F.},
title = {Galactic Bulges from Hubble Space Telescope Near-Infrared Camera Multi-Object Spectrometer Observations: The Lack of r1/4 Bulges*},
journal = {The Astrophysical Journal}
}

@ARTICLE{xue+2023,
  author={Xue, Guangdong and Chang, Qin and Wang, Jian and Zhang, Kai and Pal, Nikhil Ranjan},
  journal={IEEE Transactions on Fuzzy Systems}, 
  title={An Adaptive Neuro-Fuzzy System With Integrated Feature Selection and Rule Extraction for High-Dimensional Classification Problems}, 
  year={2023},
  volume={31},
  number={7},
  pages={2167-2181},
  doi={10.1109/TFUZZ.2022.3220950}}

@article{Freeman+2013,
    author = {Freeman, P. E. and Izbicki, R. and Lee, A. B. and Newman, J. A. and Conselice, C. J. and Koekemoer, A. M. and Lotz, J. M. and Mozena, M.},
    title = "{New image statistics for detecting disturbed galaxy morphologies at high redshift}",
    journal = {Monthly Notices of the Royal Astronomical Society},
    volume = {434},
    number = {1},
    pages = {282-295},
    year = {2013},
    month = {06},
    issn = {0035-8711},
    doi = {10.1093/mnras/stt1016},
    url = {https://doi.org/10.1093/mnras/stt1016},
    eprint = {https://academic.oup.com/mnras/article-pdf/434/1/282/18499355/stt1016.pdf},
}

@article{Lotz+2006,
doi = {10.1086/497950},
url = {https://dx.doi.org/10.1086/497950},
year = {2006},
month = {jan},
publisher = {},
volume = {636},
number = {2},
pages = {592},
author = {Jennifer M. Lotz and Piero Madau and Mauro Giavalisco and Joel Primack and Henry C. Ferguson},
title = {The Rest-Frame Far-Ultraviolet Morphologies of Star-Forming Galaxies at z ~ 1.5 and 4},
journal = {The Astrophysical Journal}
}

@InProceedings{Masci+2011,
author="Masci, Jonathan
and Meier, Ueli
and Cire{\c{s}}an, Dan
and Schmidhuber, J{\"u}rgen",
editor="Honkela, Timo
and Duch, W{\l}odzis{\l}aw
and Girolami, Mark
and Kaski, Samuel",
title="Stacked Convolutional Auto-Encoders for Hierarchical Feature Extraction",
booktitle="Artificial Neural Networks and Machine Learning -- ICANN 2011",
year="2011",
publisher="Springer Berlin Heidelberg",
address="Berlin, Heidelberg",
pages="52--59",
isbn="978-3-642-21735-7"
}

@article{Dai+2023,
doi = {10.3847/1538-4365/ace69e},
url = {https://dx.doi.org/10.3847/1538-4365/ace69e},
year = {2023},
month = {sep},
publisher = {The American Astronomical Society},
volume = {268},
number = {1},
pages = {34},
author = {Yao Dai and Jun Xu and Jie Song and Guanwen Fang and Chichun Zhou and Shuo Ba and Yizhou Gu and Zesen Lin and Xu Kong},
title = {The Classification of Galaxy Morphology in the H Band of the COSMOS-DASH Field: A Combination-based Machine-learning Clustering Model},
journal = {The Astrophysical Journal Supplement Series}
}

@article{Zhang+1996,
author = {Zhang, Tian and Ramakrishnan, Raghu and Livny, Miron},
title = {BIRCH: an efficient data clustering method for very large databases},
year = {1996},
issue_date = {June 1996},
publisher = {Association for Computing Machinery},
address = {New York, NY, USA},
volume = {25},
number = {2},
issn = {0163-5808},
url = {https://doi.org/10.1145/235968.233324},
doi = {10.1145/235968.233324},
journal = {SIGMOD Rec.},
month = {jun},
pages = {103–114},
numpages = {12}
}

@article{Murtagh+1983,
    author = {Murtagh, F.},
    title = "{A Survey of Recent Advances in Hierarchical Clustering Algorithms}",
    journal = {The Computer Journal},
    volume = {26},
    number = {4},
    pages = {354-359},
    year = {1983},
    month = {11},
    issn = {0010-4620},
    doi = {10.1093/comjnl/26.4.354},
    url = {https://doi.org/10.1093/comjnl/26.4.354},
    eprint = {https://academic.oup.com/comjnl/article-pdf/26/4/354/1072603/26-4-354.pdf},
}

@article{Murtagh+2014,
   title={Ward’s Hierarchical Agglomerative Clustering Method: Which Algorithms Implement Ward’s Criterion?},
   volume={31},
   ISSN={1432-1343},
   url={http://dx.doi.org/10.1007/s00357-014-9161-z},
   DOI={10.1007/s00357-014-9161-z},
   number={3},
   journal={Journal of Classification},
   publisher={Springer Science and Business Media LLC},
   author={Murtagh, Fionn and Legendre, Pierre},
   year={2014},
   month=oct, pages={274–295} }

@article{Hartigan+1979,
 ISSN = {00359254, 14679876},
 URL = {http://www.jstor.org/stable/2346830},
 author = {J. A. Hartigan and M. A. Wong},
 journal = {Journal of the Royal Statistical Society. Series C (Applied Statistics)},
 number = {1},
 pages = {100--108},
 publisher = {[Wiley, Royal Statistical Society]},
 title = {Algorithm AS 136: A K-Means Clustering Algorithm},
 urldate = {2024-03-13},
 volume = {28},
 year = {1979}
}

@misc{liu+2023,
  author  = {Liu, Zhaocong and Zhang, Fa and Cheng, Lin and Deng, Huanxi and Yang, Xiaoyan and Zhang, Zhenyu and Zhou, Chi-Chun},
  title   = {Simple But Effective Unsupervised Classification for Specified Domain Images: A Case Study on Fungi Images},
  year    = {2023},
  howpublished = {\url{https://ssrn.com/abstract=4673082}},
  doi     = {10.2139/ssrn.4673082}
}

@article{Gu+2018,
doi = {10.3847/1538-4357/aaad0b},
url = {https://dx.doi.org/10.3847/1538-4357/aaad0b},
year = {2018},
month = {feb},
publisher = {The American Astronomical Society},
volume = {855},
number = {1},
pages = {10},
author = {Yizhou Gu and Guanwen Fang and Qirong Yuan and Zhenyi Cai and Tao Wang},
title = {The Morphological Evolution, AGN Fractions, Dust Content, Environments, and Downsizing of Massive Green Valley Galaxies at 0.5 &lt; z &lt; 2.5 in 3D-HST/CANDELS},
journal = {The Astrophysical Journal}
}

@article{Barden+2012,
  title={GALAPAGOS: From Pixels to Parameters},
  author={Marco Barden and Boris Haussler and Chien Y. Peng and D. H. Mcintosh and Yicheng Guo},
  journal={Monthly Notices of the Royal Astronomical Society},
  year={2012},
  volume={422},
  pages={449-468},
  url={https://api.semanticscholar.org/CorpusID:119117461}
}

@article{Koekemoer+2007,
doi = {10.1086/520086},
url = {https://dx.doi.org/10.1086/520086},
year = {2007},
month = {sep},
publisher = {},
volume = {172},
number = {1},
pages = {196},
author = {A. M. Koekemoer and H. Aussel and D. Calzetti and P. Capak and M. Giavalisco and J.-P. Kneib and A. Leauthaud and O. Le Fèvre and H. J. McCracken and R. Massey and B. Mobasher and J. Rhodes and N. Scoville and P. L. Shopbell},
title = {The COSMOS Survey: Hubble Space Telescope Advanced Camera for Surveys Observations and Data Processing*},
journal = {The Astrophysical Journal Supplement Series}
}

@INPROCEEDINGS{Koekemoer+2003,
       author = {{Koekemoer}, A.~M. and {Fruchter}, A.~S. and {Hook}, R.~N. and {Hack}, W.},
        title = "{MultiDrizzle: An Integrated Pyraf Script for Registering, Cleaning and Combining Images}",
    booktitle = {HST Calibration Workshop : Hubble after the Installation of the ACS and the NICMOS Cooling System},
         year = 2003,
        month = jan,
        pages = {337},
       adsurl = {https://ui.adsabs.harvard.edu/abs/2003hstc.conf..337K}
}

@article{Song+2024,
doi = {10.3847/1538-4365/ad434f},
url = {https://dx.doi.org/10.3847/1538-4365/ad434f},
year = {2024},
month = {jun},
publisher = {The American Astronomical Society},
volume = {272},
number = {2},
pages = {42},
author = {Jie Song and GuanWen Fang and Shuo Ba and Zesen Lin and Yizhou Gu and Chichun Zhou and Tao Wang and Cai-Na Hao and Guilin Liu and Hongxin Zhang and Yao Yao and Xu Kong},
title = {USmorph: An Updated Framework of Automatic Classification of Galaxy Morphologies and Its Application to Galaxies in the COSMOS Field},
journal = {The Astrophysical Journal Supplement Series}
}

@ARTICLE{haubler+2022,
       author = {{H{\"a}u{\ss}ler}, Boris and {Vika}, Marina and {Bamford}, Steven P. and {Johnston}, Evelyn J. and {Brough}, Sarah and {Casura}, Sarah and {Holwerda}, Benne W. and {Kelvin}, Lee S. and {Popescu}, Cristina},
        title = "{GALAPAGOS-2/GALFITM/GAMA - Multi-wavelength measurement of galaxy structure: Separating the properties of spheroid and disk components in modern surveys}",
      journal = {\aap},
         year = 2022,
        month = aug,
       volume = {664},
          eid = {A92},
        pages = {A92},
          doi = {10.1051/0004-6361/202142935},
archivePrefix = {arXiv},
       eprint = {2204.05907},
 primaryClass = {astro-ph.GA},
       adsurl = {https://ui.adsabs.harvard.edu/abs/2022A&A...664A..92H}
}

@ARTICLE{li+2019,
       author = {{Li}, Zhiyuan and {Arora}, Sanjeev},
        title = "{An Exponential Learning Rate Schedule for Deep Learning}",
      journal = {arXiv e-prints},
         year = 2019,
        month = oct,
          eid = {arXiv:1910.07454},
        pages = {arXiv:1910.07454},
          doi = {10.48550/arXiv.1910.07454},
archivePrefix = {arXiv},
       eprint = {1910.07454},
 primaryClass = {cs.LG},
       adsurl = {https://ui.adsabs.harvard.edu/abs/2019arXiv191007454L}
}

@book{lehmann+2006,
  title={Theory of point estimation},
  author={Lehmann, Erich L and Casella, George},
  year={2006},
  publisher={Springer Science \& Business Media}
}

@ARTICLE{Agarap+2018,
       author = {{Agarap}, Abien Fred},
        title = "{Deep Learning using Rectified Linear Units (ReLU)}",
      journal = {arXiv e-prints},
         year = 2018,
        month = mar,
          eid = {arXiv:1803.08375},
        pages = {arXiv:1803.08375},
          doi = {10.48550/arXiv.1803.08375},
archivePrefix = {arXiv},
       eprint = {1803.08375},
 primaryClass = {cs.NE},
       adsurl = {https://ui.adsabs.harvard.edu/abs/2018arXiv180308375A}
}

@ARTICLE{Dominguez+2018,
       author = {{Dom{\'\i}nguez S{\'a}nchez}, H. and {Huertas-Company}, M. and {Bernardi}, M. and {Tuccillo}, D. and {Fischer}, J.~L.},
        title = "{Improving galaxy morphologies for SDSS with Deep Learning}",
      journal = {\mnras},
         year = 2018,
        month = feb,
       volume = {476},
       number = {3},
        pages = {3661-3676},
          doi = {10.1093/mnras/sty338},
archivePrefix = {arXiv},
       eprint = {1711.05744},
 primaryClass = {astro-ph.GA},
       adsurl = {https://ui.adsabs.harvard.edu/abs/2018MNRAS.476.3661D}
}

@book{Gonzalez+2006,
author = {Gonzalez, Rafael C. and Woods, Richard E.},
title = {Digital Image Processing (3rd Edition)},
year = {2006},
isbn = {013168728X},
publisher = {Prentice-Hall, Inc.},
address = {USA}
}

@INPROCEEDINGS{szegedy+2015,
  author={Szegedy, Christian and Wei Liu and Yangqing Jia and Sermanet, Pierre and Reed, Scott and Anguelov, Dragomir and Erhan, Dumitru and Vanhoucke, Vincent and Rabinovich, Andrew},
  booktitle={2015 IEEE Conference on Computer Vision and Pattern Recognition (CVPR)}, 
  title={Going deeper with convolutions}, 
  year={2015},
  volume={},
  number={},
  pages={1-9},
  doi={10.1109/CVPR.2015.7298594}}

@article{breiman+1996,
  author    = {Breiman, Leo},
  title     = {Bagging Predictors},
  journal   = {Machine Learning},
  year      = {1996},
  volume    = {24},
  number    = {2},
  pages     = {123--140},
  month     = aug,
  issn      = {1573-0565},
  doi       = {10.1023/A:1018054314350},
  url       = {https://doi.org/10.1023/A:1018054314350}
}

\end{document}